\documentclass[journal]{IEEEtran}
\ifCLASSINFOpdf
\else
\fi
\usepackage{amsmath}

\usepackage{cite}
\usepackage{tikz}
\usetikzlibrary{shapes.geometric}

\pgfdeclareplotmark{mystar}{
    \node[star,star point ratio=2.25,minimum size=5pt,
          inner sep=0pt,draw=blue,solid,fill=blue] {};
}
\usetikzlibrary{plotmarks}
\usetikzlibrary{calc}
\usetikzlibrary{arrows}
\usetikzlibrary{arrows,intersections}
\usetikzlibrary{arrows.meta}
\usetikzlibrary{shapes,decorations}
\usetikzlibrary{positioning}
\usepackage{pgfplots}
\tikzset{sin v source/.style={
  circle,
  draw,
  append after command={
    \pgfextra{
    \draw
      ($(\tikzlastnode.center)!0.5!(\tikzlastnode.west)$)
       arc[start angle=180,end angle=0,radius=0.425ex]
      (\tikzlastnode.center)
       arc[start angle=180,end angle=360,radius=0.425ex]
      ($(\tikzlastnode.center)!0.5!(\tikzlastnode.east)$)
    ;
    }
  },
  scale=1.5,
 }
}
\usetikzlibrary{pgfplots.groupplots, backgrounds, plotmarks, calc, spy, pgfplots.polar, shapes.geometric, arrows, external, patterns}

\tikzset{
    invisible/.style={opacity=0},
    visible on/.style={alt={#1{}{invisible}}},
    alt/.code args={<#1>#2#3}{%
      \alt<#1>{\pgfkeysalso{#2}}{\pgfkeysalso{#3}} % \pgfkeysalso doesn't change the path
    },
  }

\tikzset{sin v source/.style={
  circle,
  draw,
  append after command={
    \pgfextra{
    \draw
      ($(\tikzlastnode.center)!0.5!(\tikzlastnode.west)$)
       arc[start angle=180,end angle=0,radius=0.425ex] 
      (\tikzlastnode.center)
       arc[start angle=180,end angle=360,radius=0.425ex]
      ($(\tikzlastnode.center)!0.5!(\tikzlastnode.east)$) 
    ;
    }
  },
  scale=1.5,
 }
}

\pgfplotsset{compat=1.14}
\usepackage{graphicx}
\usepackage{booktabs}
\usepackage{amstext,epsfig,amssymb,amsbsy,amsmath}
\usepackage{cases}
\usepackage{bm}

\definecolor{lightgray}{gray}{0.7}
\newcommand*{\blue}{\textcolor{black}}

\newcommand{\mb}{\mathbf}

\newcommand*{\bl}{\textcolor{black}}

\begin{document}
% \title{Verification of Neural Network Properties \\ for Power System Security Assessment}
%\title{Formal Guarantees of Neural Network Behaviour for Power System Applications}
\title{Verification of Neural Network Behaviour:\\ Formal Guarantees for Power System Applications}

\author{Andreas~Venzke,~\IEEEmembership{Student Member,~IEEE,}
        and Spyros~Chatzivasileiadis,~\IEEEmembership{Senior Member,~IEEE}% <-this % stops a space
\thanks{A. Venzke and S. Chatzivasileiadis are with the Department of Electrical Engineering, Technical University of Denmark, 2800 Kgs. Lyngby, Denmark e-mail: \{andven, spchatz\}@elektro.dtu.dk.}
\thanks{This work  is  supported  by  the  multiDC  project,  funded  by  Innovation  Fund Denmark, Grant Agreement No. 6154-00020.}}

\maketitle

\begin{abstract}
This paper presents for the first time, to our knowledge, a framework for verifying neural network behavior in power system applications. Up to this moment, neural networks have been applied in power systems as a black box; this has presented a major barrier for their adoption in practice. Developing a rigorous framework based on mixed-integer linear programming, our methods can determine the range of inputs that neural networks classify as safe or unsafe, and are able to \blue{systematically} identify adversarial examples. Such methods have the potential to build the missing trust of power system operators on neural networks, and unlock a series of new applications in power systems. This paper presents the  framework, \blue{methods to assess and improve neural network robustness in power systems}, and addresses concerns related to scalability and accuracy. We demonstrate our methods on the IEEE 9-bus, 14-bus, and 162-bus systems, treating both \mbox{N-1} security and small-signal stability.
\end{abstract}

\IEEEpeerreviewmaketitle

\section{Introduction}
% \IEEEPARstart{T}{he} increase in fluctuating renewable energy, such as wind and solar, and the increase in electricity demand driven by the electrification of heat and transport sectors challenge traditional power system operation. 
\subsection{Motivation}
\IEEEPARstart{M}{achine} learning, such as decision trees and neural networks, has demonstrated significant potential for highly complex classification tasks including the security assessment of power systems \cite{duchesne2020recent}. However, the inability to anticipate the behavior of neural networks, which have been usually treated as a black box, has been posing a major barrier in their application in safety-critical systems, such as power systems. Recent works (e.g. \cite{goodfellow2014explaining}) have shown that neural networks that have high prediction accuracy on unseen test data can be highly vulnerable to so-called adversarial examples (small input perturbations leading to false behaviour), and that their performance is not robust. To our knowledge, the robustness of neural networks has not been systematically evaluated in power system applications before. This is the first work that develops provable formal guarantees of the behavior of neural networks in power system applications. \blue{These methods allow to evaluate the robustness and improve the interpretability of neural networks} and have the potential to build the missing trust of power system operators in neural networks, enabling their application in safety-critical operations.
\vspace{-0.1cm}
\subsection{Literature Review}
Machine learning algorithms including neural networks have been applied in power system problems for decades. For a comprehensive review, the interested reader is referred to \cite{wehenkel2012automatic,duchesne2020recent} and references therein. \blue{A recent survey in \cite{glavic2019deep} reviews applications of (deep) reinforcement learning in power systems. In the following, we focus on applications related to power system operation and security.} Recent examples include \cite{duchesne2018using} which compares different machine learning techniques for probabilistic reliability assessment and shows significant reductions in computation time compared to conventional approaches. Neural networks obtain the highest predictive accuracy. \blue{The works in \cite{Dobbe2019, Karagiannopoulos2019} rely on machine learning techniques to learn local control policies for distributed energy resources in distribution grids.} Ref.~\cite{dalal2019chance} uses machine learning to predict the result of outage scheduling under uncertainty, while in \cite{6547732} neural networks rank contingencies for security assessment. \blue{Neural networks are used in \cite{Martinez2011} to approximate the system security boundary and are then included in the security-constrained optimal power flow (OPF) problem.}

Recent developments in deep learning have sparked renewed interest for power system applications \cite{LiAlphaGo,donnot2017introducing, donnot2018fast, sun2018deep, arteaga2019deep,du2019achieving}. \blue{There are a range of applications for which deep learning holds significant promise including online security assessment, system fault diagnosis, and rolling optimization of renewable dispatch as outlined in \cite{LiAlphaGo}.} Deep neural networks are used in \cite{donnot2017introducing} to predict the line currents for different N-1 outages. By encoding the system state inside the hidden layers of the neural network, the method can be applied to unseen N-2 outages with high accuracy \cite{donnot2018fast}. The work in \cite{sun2018deep} proposes a deep autoencoder to reduce the high-dimensional input space of security assessment and increase classifier accuracy and robustness. Our work in \cite{arteaga2019deep} represents power system snapshots as images to take advantage of the advanced deep learning toolboxes for image processing. Using convolutional neural networks, the method assesses both \mbox{N-1} security and small signal stability at a fraction of the time required by conventional methods. \blue{The work in \cite{du2019achieving} uses convolutional neural networks for \mbox{N-1} contingency screening of a large number of uncertainty scenarios, and reports computational speed-ups of at least two orders of magnitude. These results highlight the potential of data-driven applications for online security assessment. The black box nature of these tools, however, presents a major obstacle towards their application in practice.}
\vspace{-0.1cm}
\subsection{Main Contributions}
\blue{In power system applications, to the best of our knowledge, the robustness of learning approaches based on neural networks has so far not been systematically investigated. Up to now, neural network performance has been evaluated by splitting the available data in a training and a test set, and assessing accuracy and other statistical metrics on the previously unseen test set data (e.g. \cite{donnot2017introducing, donnot2018fast, sun2018deep, arteaga2019deep}). Recent works (e.g. \cite{goodfellow2014explaining}) have shown that neural networks that have high prediction accuracy on unseen test data can be highly vulnerable to adversarial examples and the resulting prediction accuracy on adversarially crafted inputs is very low \cite{papernot2016limitations}. Adversarial examples also exist in power system applications as demonstrated in  \cite{chen2018machine}. This highlights the importance to develop methodologies which allow to systematically evaluate and improve the robustness of neural networks.}

\blue{In this work, we present for the first time a framework to obtain formal guarantees of neural network behaviour for power system applications, \blue{building on recent advances in the machine learning literature \cite{tjeng2017evaluating}}. Our contributions are threefold. First, we evaluate the robustness of neural networks in power system applications, through a \emph{systematic} identification of adversarial examples (small input perturbations which lead to false behaviour) or by \emph{proving} that no adversarial examples can exist for a continuous range of neural network inputs. Second, we improve the interpretability of neural networks by obtaining provable guarantees about how continuous input regions map to specific classifications. Third, using systematically identified adversarial examples, we retrain neural networks to improve robustness.}  

\bl{In the rest of this paper, we refer to verification as the process of obtaining formal guarantees for neural network behaviour. These formal guarantees are in the form of continuous input regions in which the classification of the neural network provably does not change, i.e., no adversarial examples exist.} Being able to determine the continuous \emph{range} of inputs (instead of discretized samples) that the neural network classifies as safe or unsafe makes the neural network interpretable. Accuracy is no longer a pure statistical metric but can be supported by provable guarantees of neural network behavior. This allows operators to either build trust in the neural network and use it in real-time power system operation, or decide to retrain it. Increasing the robustness, and provably verifying target properties of these algorithms is therefore \blue{a prerequisite for their application in practice}. To this end, the main contributions of our work are:

\begin{enumerate}
    \item \blue{Using mixed-integer linear programming (MILP), we present a neural network verification framework for power system applications which, for continuous ranges of inputs, can guarantee if the neural network will classify them as safe or unsafe.}
    \item \blue{We present a systematic procedure to identify adversarial examples and determine neural network input regions in which \emph{provably} no adversarial examples exist.}
    \item \blue{We improve the robustness of neural networks by re-training them on enriched training datasets that include adversarial examples identified in a systematic way.} 
    \item Formulating the verification problem as a mixed-integer linear program, we \blue{investigate, test and apply} techniques to maintain scalability; these involve bound tightening and weight sparsification.
   \item We demonstrate our methods on the IEEE 9-bus, 14-bus, and 162-bus system, treating both N-1 security and small-signal stability. \blue{For the IEEE 9-bus system, we re-train the neural network using identified adversarial examples, and show improvements both in accuracy and robustness.}
\end{enumerate}
This work is structured as follows: In Section~\ref{sec:NNarchitecture}, we describe the neural network architecture and training, in Section~\ref{sec:verificationMIP} we formulate the verification problems as  mixed-integer programs and in Section~\ref{sec:Tractability} we address tractability. In Section~\ref{sec:simresults} we define our simulation setup and present results on formal guarantees for a range of power system security classifiers.

\section{Neural Network Architecture and Training}
\label{sec:NNarchitecture}
\subsection{Neural Network Structure}
\begin{figure}
  \def\layersep{1.5cm}
\centering

\resizebox{0.85\columnwidth}{!}{%
\begin{tikzpicture}[shorten >=1pt,->,draw=black!50, node distance=\layersep]
    \tikzstyle{every pin edge}=[<-,shorten <=1pt]
    \tikzstyle{neuron}=[circle,fill=black!25,minimum size=17pt,inner sep=0pt]
    \tikzstyle{input neuron}=[neuron];
    \tikzstyle{output neuron}=[neuron];
    \tikzstyle{hidden neuron}=[neuron];
    \tikzstyle{annot} = [text width=4em, text centered]

    % Draw the input layer nodes
    \foreach \name / \y in {1,...,3}
    % This is the same as writing \foreach \name / \y in {1/1,2/2,3/3,4/4}
        \node[input neuron] (I-\name) at (0,-\y) {$x_\y$};

    % Draw the hidden layer nodes
    \foreach \name / \y in {1,...,4}
        \path[yshift=0.5cm]
            node[hidden neuron] (H1-\name) at (\layersep,-\y cm) {$z^\y_1$};
            
                % Draw the hidden layer nodes
    \foreach \name / \y in {1,...,4}
        \path[yshift=0.5cm]
            node[hidden neuron,right of=H1-\name] (H2-\name) at (\layersep,-\y cm) {$z^\y_2$};
            
                % Draw the hidden layer nodes
    \foreach \name / \y in {1,...,4}
        \path[yshift=0.5cm]
            node[hidden neuron,right of=H2-\name] (H3-\name) at (\layersep+\layersep,-\y cm) {$z^\y_3$};

    % Draw the output layer node
    \node[output neuron,right of=H3-2] (O1) {$y_1$};
    \node[output neuron,right of=H3-3] (O2) {$y_2$};
    % Connect every node in the input layer with every node in the
    % hidden layer.
    \foreach \source in {1,...,3}
        \foreach \dest in {1,...,4}
            \path (I-\source) edge (H1-\dest);
            
                \foreach \source in {1,...,4}
        \foreach \dest in {1,...,4}
            \path (H1-\source) edge (H2-\dest);
            
                \foreach \source in {1,...,4}
        \foreach \dest in {1,...,4}
            \path (H2-\source) edge (H3-\dest);

    % Connect every node in the hidden layer with the output layer
    \foreach \source in {1,...,4}
        \path (H3-\source) edge (O1);
        
            % Connect every node in the hidden layer with the output layer
    \foreach \source in {1,...,4}
        \path (H3-\source) edge (O2);

    % Annotate the layers $(W_1,b_1)$
    \node[annot,above of=H1-1, node distance=1cm] (hl1) {Hidden layer 1};
    \node[annot,above of=H1-1, node distance=0.35cm] (w1b1) {};
    \node[annot,left of=w1b1, node distance=0.85cm] (w1b1t) {$\mb{W}_1,\mb{b}_1$};
    \node[annot,above of=H2-1, node distance=1cm] (hl2) {Hidden layer 2};
        \node[annot,above of=H2-1, node distance=0.35cm] (w2b2) {};
    \node[annot,left of=w2b2, node distance=0.75cm] (w2b2t) {$\mb{W}_2,\mb{b}_2$};
    \node[annot,above of=H3-1, node distance=1cm] (hl3) {Hidden layer 3};
    \node[annot,above of=H3-1, node distance=0.35cm] (w3b3) {};
        \node[annot,left of=w3b3, node distance=0.75cm] (w3b3t) {$\mb{W}_3,\mb{b}_3$};
    \node[annot,left of=hl1] {Input layer};
    \node[annot,right of=hl3] {Output layer};
        \node[annot,right of=w3b3, node distance=0.75cm] (w4b4t) {$\mb{W}_4,\mb{b}_4$};
\end{tikzpicture}
}
\vspace{-0.3cm}
    \caption{Illustrative neural network for binary classification: Fully connected neural network with three inputs $\mb{x}$ and three hidden layers. Between each layer, a weight matrix $\mb{W}$ and bias $\mb{b}$ is applied. Each hidden layer has four neurons with non-linear ReLU activation functions. Based on the comparison of the two outputs $\mb{y}$, the input is classified in one of the two categories, i.e. $y_1 > y_2$ or $y_1 < y_2$.  }
    \label{NN_structure}
    \vspace{-0.3cm}
\end{figure}
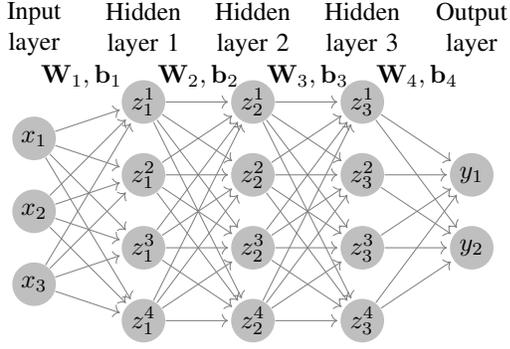

Before moving on with the formulation of the verification problem, in this section we explain the general structure of the neural networks we consider in this work \cite{Bishop_PRML}. 
An illustrative example is shown in Fig.~\ref{NN_structure}. A neural network is defined by the number $K$ of its fully-connected hidden layers, with each layer having $N_k$ number of neurons (also called nodes or hidden units), with $k = 1, ..., K$. The input vector is denoted with $\mb{x} \in \mathbb{R}^{N_0}$, and the output vector with $\mb{y} \in \mathbb{R}^{N_{K+1}}$. The input to each layer $\hat{\mb{z}}_{k+1}$ is a linear combination of the output of the previous layer, i.e. $\hat{\mb{z}}_{k+1}=\mb{W}_k\mb{z}_k+\mb{b}_k$, where $\mb{W}_k$ is a $N_{k+1} \times N_k$ weight matrix and $\mb{b}_k$ is a $N_{k+1} \times 1$ bias vector between layers $k$ and $k+1$. Each neuron in the hidden layers incorporates an activation function, $z = f(\hat{z})$, which usually applies a non-linear transformation to the scalar input. There is a range of possible activation functions, such as the sigmoid function, the hyberbolic tangent, the Rectifier Linear Unit (ReLU), and others. 
\blue{Recent advances in computational power and machine learning have made possible the successful training of deep neural networks \cite{DeepLearning_Nature}. The vast majority of such networks use ReLU as the activation function as this has been shown to accelerate their training \cite{glorot2011deep}. For the rest of this paper we will focus on ReLU as the chosen activation function. A framework for neural network verification considering general activation functions such as the sigmoid function and the hyberbolic tangent is proposed in \cite{NIPS2018_7742}. ReLU is a piecewise linear function defined as $z = \max(\hat{z},0)$. }For each of the hidden layers we have:
\begin{alignat}{2}
    \mb{z}_k & = \max(\hat{\mb{z}}_k,0) &&  \quad \forall \, k = 1, ...,K \label{ReLU_max} \\
    \hat{\mb{z}}_{k+1} &= \mb{W}_{k+1}\mb{z}_k + \mb{b}_{k+1} &&  \quad \forall \, k = 0, 1, ...,K-1 \label{NN_layer}
\end{alignat}
where $\mb{z}_0=\mb{x}$, i.e. the input vector. Throughout this work, the $\max$ operator on a vector $\hat{\mb{z}}_k \in \mathbb{R}^{N_{k}}$ is defined element-wise as $z_k^n = \max(\hat{z}_k^n,0)$ with $n = 1, ..., N_k$. The output vector is then obtained as follows:
\begin{align}
    \mb{y} & = \mb{W}_{K+1} \mb{z}_K + \mb{b}_{K+1} \label{NN_output}
\end{align}
In this work, we will focus on classification networks, that is each of the output states $y_i$ corresponds to a different class. For example, within the power systems context, each operating point $\mb{x}$ can be classified as $y_1$=\,\emph{safe} or $y_2$=\,\emph{unsafe} (binary classification). \blue{The input vector $\mb{x}$ encodes the necessary information to determine the operating point, e.g. the generation dispatch and loading in the DC optimal power flow (OPF).} Each input is classified to the category that corresponds to the largest element of the output vector $\mb{y}$. For example, if $y_1 > y_2$ then input $\mb{x}$ is safe, otherwise unsafe.  

%\vspace{-0.7cm}
\subsection{Neural Network Training}
The training of neural networks requires a dataset of labeled samples $\mb{S}$, where each sample consists of the input $\mb{x}$ and the true output classification $\bar{\mb{y}}$. The dataset $\mb{S}$ is split into a training and a testing dataset \blue{denoted with $\mb{S}_{\text{train}}$ and $\mb{S}_{\text{test}}$, respectively}. During training, the weights $\mb{W}$ and biases $\mb{b}$ are optimized using the training dataset $\mb{S}_{\text{train}}$ with respect to a defined objective function. Different objective functions (also called loss functions) exist \cite{Bishop_PRML}. In this work we use one of the most commonly employed for classification: the softmax cross entropy, which is defined as follows. First, the softmax function takes the vector of the neural network output $\mb{y}$ and transforms it to an equal-size vector of real numbers in the range between 0 and 1, the sum of which have to be equal to 1. \blue{This corresponds to a probability distribution of the output of the neural network belonging to a certain class.} The probability $p_i$ of the input belonging to the different classes $i \in N_{K+1}$ is defined as:
\begin{align}
    p_i = \tfrac{e^{y_i}}{\sum_{j=1}^{N_{K+1}} e^{y_j}}, \quad \forall i \in N_{K+1} \label{p_i}
\end{align}
\blue{Note that $e^{\{.\}}$ refers to the exponential function here. We define the softmax cross entropy function as:}
\begin{align}
 \blue{f(\mb{\bar{y}},\mb{p}) = - \sum_{i = 1}^{N_{K+1}} \bar{y}_i \log (p_i) \label{cross_entr}} %\label{cross}
\end{align}
\blue{This objective function can be understood as the squared error of the distance from the true classification $\bar{\mb{y}}$ to the probability distribution over the classes $\mb{p}$ predicted by the neural network. This formulation has theoretical and practical advantages over penalizing the squared error directly \cite{kline2005revisiting}. During training, we solve the following optimization problem:
\begin{align}
    \min_{\mb{W},\mb{b},\mb{x},\mb{y},\mb{\bar{y}},\mb{z},\hat{\mb{z}},\mb{p}} \, \, & f(\mb{\bar{y}},\mb{p}) \label{NN_train1} \\
    \text{s.t.} \quad & \eqref{ReLU_max}, \eqref{NN_layer}, \eqref{NN_output}, \eqref{p_i} \text{ and } \eqref{cross_entr} \label{NN_train2} %\\
\end{align}
The training follows these steps: First, we initialize the weights $\mb{W}$ and biases $\mb{b}$ randomly. Second, we substitute the variables $\mb{x}$ and $\mb{\bar{y}}$ with the samples from the training dataset $\mb{S}_{\text{train}}$. Based on the current value of weights $\mb{W}$ and biases $\mb{b}$, we compute the corresponding neural network prediction $\mb{y}$, and the loss function \eqref{cross_entr}. Third, using back-propagation, we compute the gradients of the loss function with respect to weights $\mb{W}$  and biases $\mb{b}$ and update these using stochastic gradient descent (in our simulation studies we use the Adam optimizer \cite{tensorflow2015-whitepaper}). Then, we return to the second step, and repeat this procedure until a defined number of training iterations, also called epochs, is reached. After the training has terminated, we evaluate the neural network performance by calculating its accuracy on the unseen test dataset $\mb{S}_{\text{test}}$. This gives us a measure of the generalization capability of the neural network. For a detailed explanation of neural network training please refer to e.g. \cite{glorot2010understanding}.}
\vspace{-0.15cm}
\section{Verification as a Mixed-Integer Program}
\label{sec:verificationMIP}
As the accuracy on the test dataset is not sufficient to provide provable guarantees of neural network behavior, in the following, we formulate verification problems that allow us to verify target properties of neural networks and identify adversarial examples in a rigorous manner. \bl{To this end, we first reformulate trained neural networks as mixed-integer programs. Then, we introduce verification problems to a)~prove the absence of adversarial examples, b)~evaluate the neural network robustness, and c)~compute largest input regions with same classification. Finally, we discuss how to include additional constraints on the input to the neural network and how to extend the framework to regression problems.}

\vspace{-0.15cm}
\subsection{Reformulation of ReLU as Mixed-Integer Program (MIP)}
 First, we reformulate the ReLU function $\mb{z}_k = \max(\hat{\mb{z}}_k,0)$, shown in \eqref{ReLU_max}, using binary variables $\mb{r}_k \in  \{0,1\}^{N_k}$, following the work in  \cite{tjeng2017evaluating}. 
%\textcolor{red}{again the equations below are for a vector or a scalar. If scalar, it must be $z_k^n$}
For all $k = 1, ...,K$, it holds:
\begin{subnumcases}{\mb{z}_k = \max(\hat{\mb{z}}_k,0)\Rightarrow}
\mb{z}_k  \leq \hat{\mb{z}}_k - \hat{\mb{z}}^{\text{min}}_k  (1-\mb{r}_k) \label{ReLU1} \\ 
\mb{z}_k  \geq \hat{\mb{z}}_k \label{ReLU2}   \\
\mb{z}_k  \leq \hat{\mb{z}}^{\text{max}}_k \mb{r}_k  \label{ReLU3}  \\
\mb{z}_k   \geq \mb{0}  \label{ReLU4}  \\
\mb{r}_k \in \{0,1\}^{N_k} \label{ReLUe}
\end{subnumcases}
The entries of the binary vector $\mb{r}_k$ signal whether the corresponding ReLU unit in layer $k$ is active or not. Note that the operations in \eqref{ReLU1}--\eqref{ReLU4} are performed element-wise. \blue{For a ReLU unit $n$, if $r_k^n=0$, the ReLU is inactive and $z_{k}^n$ is constrained to be zero through \eqref{ReLU3} and \eqref{ReLU4} given that the expression $\mb{0} \leq \hat{\mb{z}}_k - \hat{\mb{z}}^{\text{min}}_k$ is true}. \blue{Conversely, if $r_k^n$ is 1, then $z_{k}^n$ is constrained to be equal to $\hat{z}_{k}^n$ through \eqref{ReLU1} and \eqref{ReLU2} given that the expression $\hat{\mb{z}}_k \leq \hat{\mb{z}}^{\text{max}}_k$ is true}. \blue{To ensure that both expressions are always true, the bounds on the ReLU output $\hat{\mb{z}}^{\text{min}}$ and $\hat{\mb{z}}^{\text{max}}$ have to be chosen sufficiently large to not be binding, but as low as possible to provide tight bounds in the branch-and-bound algorithm during the solution of the MIP.} In Section~\ref{subsec:IA}, we present a possible approach to tighten these bounds using interval arithmetic (IA). 
%\vspace{-0.3cm}
\subsection{Formulating Verification Problems}
Using \eqref{ReLU1}-\eqref{ReLUe}, we can now verify neural network properties by formulating mixed-integer linear programs (MILPs). Without loss of generality, in this paper we assume that the neural network classifies the output in only two categories: $y_1$ and $y_2$ (e.g. safe and unsafe). \blue{Note that in case the neural network outputs are of the same magnitude ($y_1 = y_2$), we classify the input as `unsafe'; this avoids the risk of classifying a true `unsafe' operating point as `safe' in those cases.}
\paragraph{Verifying against adversarial examples}
\label{sec:verifyagainstadverarial}
Assume a given input $\mb{x}_{\text{ref}}$ is classified as $y_1$, i.e. for the neural network output holds $y_1 > y_2$. Adversarial examples are instances close to the original input $\mb{x}_{\text{ref}}$ that result to a different (wrong) classification \cite{goodfellow2014explaining}. Imagine for example an image of $2000 \times 1000$ pixels showing a green traffic light. An adversarial example exists if by changing just 1 pixel at one corner of the image, the neural network would classify it as a red light instead of the initial classification as green (e.g. see Fig. 16 of \cite{wu2019game}). Machine learning literature reports on a wide range of such examples and methods for identifying them, as they can be detrimental for safety-critical application (such as autonomous vehicles). Due to the nature of this problem, however, most of these methods rely on heuristics. Verification can be a very helpful tool towards this objective as it can help us discard areas around given inputs by providing guarantees that no adversarial example exists \cite{tjeng2017evaluating}. 

Turning back to our problem, assume that the system operator knows that within distance $\epsilon$ from the operating point $\mb{x}_{\text{ref}}$ the system remains safe ($y_1 > y_2$). If we can guarantee that the neural network will output indeed a safe classification for any input $\| \mb{x} - \mb{x}_{\text{ref}}\| \leq \epsilon$, then we can provide the operator with the guarantees required in order to deploy methods based on neural networks for real-time power system applications. To this end, we can solve the following mixed-integer program:
\begin{subequations}
\label{Adv_OPT} 
\begin{alignat}{2}
\min_{\mb{x},\hat{\mb{z}},\mb{z},\mb{y}} \quad & y_1 - y_2 && \label{obj_VF1}\\ 
\text{s.t.} \quad & \eqref{NN_layer}, \eqref{NN_output},  \eqref{ReLU1}-\eqref{ReLUe} && \\
 & \| \mb{x} - \mb{x}_{\text{ref}}\|_{*} \leq \epsilon && \label{Norm_eps}
\end{alignat}
\end{subequations}
If the resulting objective function value is strictly positive (and assuming zero optimality gap), then $y_1 > y_2$ for all $\| \mb{x} - \mb{x}_{\text{ref}}\|_{*} \leq \epsilon$, and we can guarantee that no adversarial input exists within distance $\epsilon$ from the given point.
The norm in \eqref{Norm_eps} can be chosen to be e.g. $\infty$-norm, $1$-norm or $2$-norm. \blue{In the following, we focus on the $\infty$-norm, which allows us to formulate the optimization problem \eqref{Adv_OPT} as MILP. In addition, the $\infty$-norm has a natural interpretation of allowing each input dimension in magnitude to vary at most by $\epsilon$.} \bl{By considering both the 1- and $\infty$-norm in \eqref{Norm_eps}, adversarial robustness with respect to all $l_p$-norms with $p\geq2$ can be achieved \cite{croce2019provable}. }
 % If we solve the MILP \eqref{obj_VF1} -- \eqref{Norm_eps} to zero optimality gap and the resulting objective value is strictly positive, then $y_1 > y_2$ for all $\| \mb{x} - \mb{x}_{\text{ref}}\|_{*} \leq \epsilon$. By that, we have obtained a guarantee that no input $x$ exists within a box of size of $2 \epsilon \times 2 \epsilon$ around $\mb{x}_{\text{ref}}$ that changes the initial classification. %, $l_p$ \textcolor{red}{infinity and 2 belong to Lp}
 Conversely, in case input $\mb{x}_{\text{ref}}$ was originally classified as $y_2$, we minimize $y_2 - y_1$ in objective function \eqref{obj_VF1} instead.  \blue{Note that we model $\mb{z}$, $\hat{\mb{z}}$ and $\mb{y}$ as optimization variables in the mixed-integer program in \eqref{Adv_OPT}, as we are searching for an input $\mb{x}$ with output $\mb{y}$ that changes the classification. For fixed input $\mb{x}$ these optimization variables are then uniquely determined.}
\paragraph{Adversarial robustness} Instead of solving \eqref{Adv_OPT} for a single $\epsilon$ and a single operating point $\mb{x_{ref}}$, we can solve a series of optimization problems \eqref{Adv_OPT} for different values of $\epsilon$ and different operating points $\mb{x_{ref}}$ and assess the adversarial accuracy as a measure of neural network robustness. The adversarial accuracy is defined as the share of samples that do not change classification from the correct ground-truth classification within distance $\epsilon$ from the given input $\mb{x_{ref}}$. In our simulation studies, we use all samples in the training data or unseen test data set to evaluate the adversarial accuracy. As such, the adversarial accuracy for a distance measure of zero ($\epsilon = 0$) is equal to the share of correctly predicted samples, i.e. the prediction accuracy. The adversarial accuracy can be used as an indicator for the robustness/brittleness of the neural network: low adversarial accuracy for very small $\epsilon$ is in most cases an indicator of poor neural network performance \cite{madry2017towards}. 

\bl{Furthermore, utilizing this methodology, we can systematically identify adversarial examples to evaluate the adversarial robustness and re-train neural networks to improve their adversarial robustness. First, for a range of different values of $\epsilon$ and using either training or test dataset, we solve the optimization problems in \eqref{Adv_OPT}. For all samples $\mb{x_{\text{ref}}}$ for which an input perturbation exists within a distance of $\epsilon$ that does change the classification, we need to assess whether this change occurs as the sample is located at the power system security boundary (and the input perturbation places the adversarial input across the boundary) or if it is in fact an adversarial example and the change in classification indicates an incorrect boundary prediction by the neural network. This can be achieved by computing the ground-truth classification $\mb{\bar{y}}$ for the perturbed input, e.g., by evaluating the system security and stability using conventional methods for the identified potential adversarial inputs, and comparing it to the neural network prediction. The share of identified adversarial examples (i.e. false classification changes) serves as measure of adversarial robustness.}

\bl{In our simulation studies in Section~\ref{sec:simresults}, we compute the adversarial accuracy and identify adversarial examples for several illustrative test cases using the proposed methodology. In case the neural network robustness is not satisfactory, we can use the aforementioned procedure to systematically identify adversarial examples from the training dataset, add these to the training dataset and re-train the neural network by optimizing \eqref{NN_train1}--\eqref{NN_train2} using the enriched training dataset to improve robustness. Other directions for improving the performance of the neural network include to (i) use a dataset creation method that provides a more detailed description of the security boundary e.g. using the algorithm in \cite{thams2019efficient}, and (ii) re-train the neural networks to be adversarially robust by modifying the objective function \eqref{cross_entr} as outlined e.g. in \cite{madry2017towards}.}
\paragraph{Computing largest regions with same classification} 
\begin{figure}
% Define block styles
\tikzstyle{decision} = [diamond, draw, fill=gray!20,
  text centered, node distance=3cm, inner sep=0pt]
\tikzstyle{block} = [rectangle, draw, fill=gray!20, 
    text centered, rounded corners, minimum height=0em]
\tikzstyle{line} = [draw, -latex']

    \begin{center}
\begin{tikzpicture}[node distance = 1.5cm, auto]
    % Place nodes
    \node [block] (Dataset) {$\begin{matrix} \text{Create training and test dataset} \\ \mb{S_{\text{train}}} \text{ and } \mb{S_{\text{test}}}\end{matrix}$};
    \node [block, below of=Dataset] (Training) {$\begin{matrix} \text{Standard neural network training procedure:} \\ \text{optimize \eqref{NN_train1} -- \eqref{NN_train2} using } \mb{S_{\text{train}}} \end{matrix}$};
        \node [block, below of=Training] (Confusion) {$\begin{matrix} \text{Evaluate confusion matrix and} \\ \text{accuracy using } \mb{S_{\text{train}}} \text{ and } \mb{S_{\text{test}}} \end{matrix}$};
                \node [block, below of=Confusion,node distance = 2cm] (Adversarial) {$\begin{matrix} \text{Use }\mb{S_{\text{test}}}\text{; evaluate adversarial accuracy}   \\ \text{by solving \eqref{Adv_OPT} for different } \epsilon \text{ and } \mb{x}_{\text{ref}} \text{;}\\  \text{ to identify adversarial examples compute} \\ \text{ ground-truth classification (Step A) }  \end{matrix}$};
                 \node [decision,below of=Adversarial,node distance = 2.5cm] (Decision){$\begin{matrix} \text{Is robustness} \\ \text{satisfactory?} \end{matrix}$};
                 \node [block,right of=Decision,node distance = 2.5cm] (end){end};
                 \node [block,below of=Decision,node distance = 2.5cm] (Adversarial_train){$\begin{matrix} \text{Use } \mb{S_{\text{train}}} \text{ and repeat previous step A;}\\  \text{add identified adversarial examples to } \mb{S_{\text{train}}} \end{matrix}$};
              %   \node [block,right of =tmp,node distance = 2cm] (Adversarial_train) {$\begin{matrix} \text{Use } \mb{S_{\text{train}}} \text{ and add identified} \\  \text{adversarial examples  to } \mb{S_{\text{train}}}\end{matrix}$};
                      %           \node [block,left of =tmp,node distance = 2cm] (Adversarial_test) {$\begin{matrix} \text{Use } \mb{S_{\text{test}}} \text{ to } \\  \text{evaluate robustness} \end{matrix}$};
                                                              \node [block,below of =Adversarial_train,node distance = 1.75cm] (Guarantees) {$\begin{matrix} \text{Solve \eqref{Ver_OPT} to determine largest input} \\  \text{regions with guaranteed classification} \end{matrix}$};
                                                 \path [line] (Dataset) -- (Training);
                                                       \node [right of=Training] (tmp2){};
                                                       \node [left of=Confusion,node distance = 3.5cm,inner sep=0pt,minimum size=0pt] (tmp4){};
                                                        \node [right of=tmp2,circle,inner sep=0pt,minimum size=0pt,node distance = 3cm] (tmp3){};
                \path [line] (Dataset) -- (Training);
    \path [line] (Training) -- (Confusion);
        \path [line] (Confusion) -- (Adversarial);
           \path [line] (Adversarial) -- (Decision);
           \path [line] (Decision) -- node [near start] {no} (Adversarial_train);
              \path [line] (Decision) -- node [near start] {yes} (end);
          %        \path [line] (Adversarial) -- (Adversarial_test);
                       %\path [line] (Adversarial_test) -- (Guarantees);
                          \path [line] (tmp3) -- (Training);
                          \path [draw] (Confusion) -- (tmp4);
                          \path [line] (tmp4) |- (Guarantees);
                          \path [draw] (Adversarial_train) -| (tmp3);
                        %  \path [draw] (tmp2) |- (Training);
 %   \node [block, below of=identify] (evaluate) {evaluate candidate models};
 %   \node [block, left of=evaluate, node distance=3cm] (update) {update model};
 %   \node [decision, below of=evaluate] (decide) {is best candidate better?};
 %   \node [block, below of=decide, node distance=3cm] (stop) {stop};
    % Draw edges
 %   \path [line] (init) -- (identify); %   \path [line] (identify) -- (evaluate);
 %   \path [line] (evaluate) -- (decide);
 %   \path [line] (decide) -| node [near start] {yes} (update);
 %   \path [line] (update) |- (identify);
 %   \path [line] (decide) -- node {no}(stop);
\end{tikzpicture}
\end{center}
\vspace{-0.3cm}
\caption{\bl{Flowchart illustrating the methodology: First, we create datasets which are split into training and test set. Using the training set only, we train the neural network. Following standard procedure, we evaluate the neural network performance with the confusion matrix. Then, using the mixed-integer linear reformulation of the trained neural network and the test dataset, we evaluate the adversarial accuracy and identify adversarial examples. If the neural network robustness is not satisfactory, we repeat this step with the training test set, add the identified adversarial examples to the training set and re-train the neural network. Note that we cannot use the test set for this step, as the test set information should not be used in the training process; instead the test set should only be used to evaluate the generalization capability on unseen data. Finally, as shown in the last block as a separate stream, with our methods we are also able to determine input regions around selected input samples with guaranteed classification to provide formal guarantees for neural network behaviour.}}
\label{flowchart}
\vspace{-0.5cm}
\end{figure}
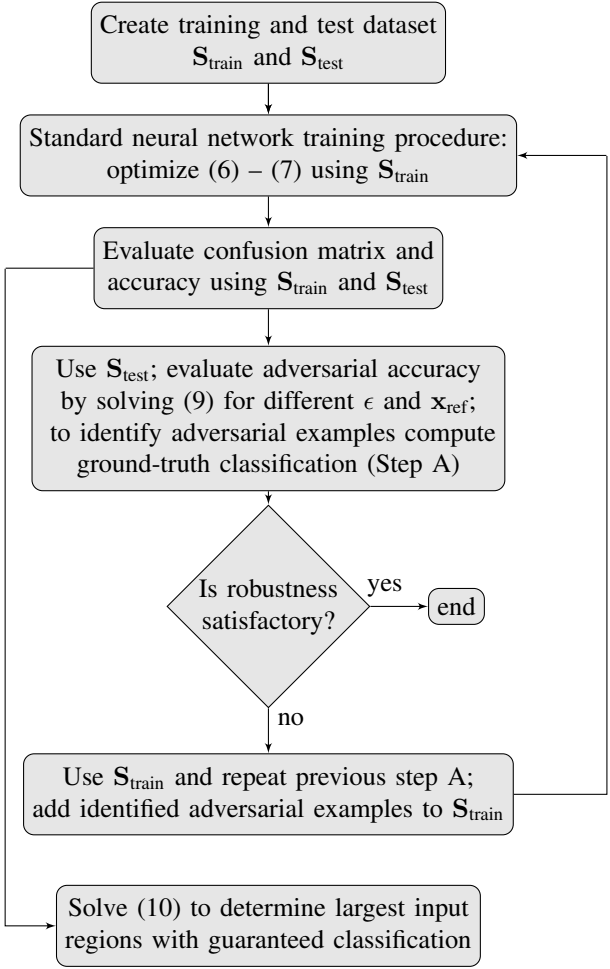
To establish trust of neural networks among power system operators it is crucial to be able to determine the range of inputs a neural network will classify as safe or unsafe. The formulation \eqref{Ver_OPT} does that by computing the maximum input range around a given input $\mb{x_{ref}}$ for which the classification will not change. Note that we achieve this by computing the \emph{minimum} distance to a sample which \emph{does} change the classification: 
\begin{subequations}
\label{Ver_OPT}
\begin{alignat}{2}
\min_{\mb{x},\hat{\mb{z}},\mb{z},\mb{y},\epsilon} \quad & \epsilon && \label{obj_VF2}\\ 
\text{s.t.} \quad & \eqref{NN_layer}, \eqref{NN_output},  \eqref{ReLU1}-\eqref{ReLUe} && \\
 & \| \mb{x} - \mb{x}_{\text{ref}}\|_{*} \leq \epsilon &&  \label{Norm_eps2} \\
 & y_2 \geq y_1 &&  \label{Fix_clas}
\end{alignat}
\end{subequations}
Here again we select the $\infty$-norm in \eqref{Norm_eps2}, turning \eqref{Ver_OPT} to a MILP. If input $\mb{x}_{\text{ref}}$ is classified as $y_2$, then we replace \eqref{Fix_clas} with $y_1 \geq y_2$ instead.  Solving verification problem \eqref{Ver_OPT} enables us to provide the operator with guarantees that e.g. all system states where Generator 1 produces between 50~MW and 100~MW and Generator 2 produces between 30~MW and 80~MW, will be classified by the neural network as safe. Thus, the neural network is no longer a black box, and it is up to the operator to determine if its results are accurate enough. If they are not, the operator can retrain the neural network with adjusted parameters to increase the accuracy for regions  the operator is mostly interested. Alternatively,  the operator can follow a risk-based approach attaching a certain confidence/risk level to the neural network results, or the operator can choose to use the neural network for a limited range of inputs where its results are provably 100\% accurate. \bl{Our overall methodology is illustrated in a flowchart in Figure~\ref{flowchart}.}
\paragraph{Input constraints} In all verification problems, we can add additional constraints characterizing the admissible region for the input vector $\mb{x}$ and, thus, can obtain larger regions with the same classification. Given that in neural network training it is standard practice to normalize the input between $\mb{0}$ and $\mb{1}$, we add constraint \eqref{admis_x} to both formulations in \eqref{Adv_OPT} and \eqref{Ver_OPT}:
\begin{align}
    \mb{0} \leq \mb{x} \leq \mb{1}  \label{admis_x}
\end{align}
These limits correspond for example to the minimum and maximum generator limits if the generator limits are included as part of the input vector. In Section~\ref{sec:9busresults}, we will show how including the DC power balance \eqref{DC_PB} as additional constraint on the input allows to further improve the obtained bounds in \eqref{obj_VF2}.
\paragraph{Regression problems}
\blue{ Note that the proposed framework can be extended to regression problems. As an example, neural networks can be used to predict the security margin of the power systems (e.g. the damping ratio of the least damped mode in small signal stability analysis). Then, we can solve verification problems of similar structure as \eqref{Ver_OPT} and \eqref{Adv_OPT} to determine regions in which the neural network prediction is guaranteed to deviate less than a predefined constant from the actual value.
At the same time, for a defined region, we can compute the minimum and maximum security margin predicted. This allows us to analyse the robustness of regression neural networks.}

\section{Improving Tractability of Verification Problems}
\label{sec:Tractability}
By examining the complexity of the MILPs presented in Section~\ref{Verification}, we make two observations. First, from \eqref{ReLU1} -- \eqref{ReLUe} it becomes apparent that the number of required binaries in the MILP is equal to the number of ReLU units in the neural network. Second, the weight matrices $\mb{W}$ are dense as the neural network layers are fully connected. As neural networks can potentially have several hidden layers with a large number of hidden neurons, improving the computational tractability of the MILP problems is necessary. To this end, we apply two different methods from \cite{tjeng2017evaluating} and \cite{zhu2017prune}: tightening the bounds on the ReLU output, and sparsifying the weight matrices.
\subsubsection{ReLU bound tightening} \label{subsec:IA} In the reformulation of the maximum operator in \eqref{ReLU1} and \eqref{ReLU3}, we introduced upper and lower bounds $\hat{\mb{z}}^{\text{max}}$ and $\hat{\mb{z}}^{\text{min}}$, respectively. Without any specific knowledge about the bounds, these have to be set to a large value in order not to become binding. A computationally cheap approach to tighten the ReLU bounds is through interval arithmetic (IA) \cite{tjeng2017evaluating}. We propagate the initial bounds on the input $\mb{z}_0^{\text{min}} = \mb{x}^{\text{min}}$, $\mb{z}_0^{\text{max}} = \mb{x}^{\text{max}}$  through each layer to obtain individual bounds on each ReLU in layer $k = 1,...,K$:
\begin{subequations}
\label{IA}
\begin{align}
    \hat{\mb{z}}_{k}^{\text{max}} =  \mb{W}_{k-1}^+ \hat{\mb{z}}_{k-1}^{\text{max},+}  +  \mb{W}_{k-1}^- \hat{\mb{z}}_{k-1}^{\text{min},+} + \mb{b}_{k-1}&  \label{IA_1} \\
    \hat{\mb{z}}_{k}^{\text{min}} =    \mb{W}_{k-1}^+ \hat{\mb{z}}_{k-1}^{\text{min},+} + \mb{W}_{k-1}^- \hat{\mb{z}}_{k-1}^{\text{max},+} + \mb{b}_{k-1}& \label{IA_2}
\end{align}
\end{subequations}
 The max and min-operator are denoted in compact form as: $\mb{x}^+ =  \max(\mb{x},0)$ and $\mb{x}^- = \min(\mb{x},0)$. \blue{For example, in our simulation studies, we restrict the input $\mb{x}$ to be between $\mb{x}^{\text{max}} = \mb{0}$ and $\mb{x}^{\text{min}} = \mb{1}$. Then, the bounds on the first layer evaluate to $\hat{\mb{z}}_{1}^{\text{max}} = \mb{W}_{0}^+  \mb{1}  +  \mb{b}_{0}$ and $\hat{\mb{z}}_{1}^{\text{min}}  = \mb{W}_{0}^-  \mb{1} + \mb{b}_{0}$. The bounds for the remaining layers can be obtained by applying \eqref{IA} sequentially.} Methods to compute tighter bounds also exist, e.g. by solving an LP relaxation while minimizing or maximizing the ReLU bounds \cite{Dvijotham18}, but this is out of scope of this paper. 
\subsubsection{Training Neural Networks for Easier Verifiability} \label{sec:Easier_Ver} A second possible approach to increase the computational tractability of the verification problems is to (re-)train the neural network with the additional goal to sparsify the weight matrices $\mb{W}$. Here, we rely on an automated gradual pruning algorithm proposed in \cite{zhu2017prune}. Starting from 0\% sparsity, where all weight matrices $\mb{W}$ are non-zero, a defined share of the weight entries are set to zero. The weight entries selected are those with the smallest absolute magnitude. Subsequently, the neural network is re-trained for the updated sparsified structure. This procedure is repeated until a certain sparsity target is achieved. 
There are two important observations: First, through sparsification the classification accuracy can decrease, as less degrees of freedom are available during training. Second, larger sparsified networks can achieve better performance than smaller dense  networks \cite{zhu2017prune}. As a result, by forming slightly larger neural networks we can maintain the required accuracy while achieving sparsity. As we will see in Section~\ref{sec:simresults}, sparsification maintains a high classification accuracy and, thus, a significant computational speed-up is achieved when solving the MILPs. As a further benefit of sparsification, the interpretability of the neural network increases, as the only neuron connections kept are the ones that have the strongest contribution to the classification. Computational tractability can further increase by pruning ReLUs during training, i.e. fixing them to be either active or inactive, in order to eliminate the corresponding binaries in the MILP \cite{xiao2018training}. This approach along with the LP relaxation for bound tightening will be object of our future work.

\section{Simulation and Results}
\label{sec:simresults}

\subsection{Simulation Setup}
The goal of the following simulation studies is to illustrate the proposed methodology for neural networks which classify operating points as `safe' or `unsafe' with respect to different power system security criteria. The neural network input $\mb{x}$ encodes variables such as active generator dispatch and uniquely determines an operating point. The neural network output $\mb{y}$ corresponds to the two possible classifications:  `safe' or `unsafe' with respect to the specified security criteria.

\bl{In the following, we will present four case studies. The first two case studies use a 9-bus and 162-bus system, respectively. The security criteria is feasibility to the N-1 security constrained DC optimality power flow (OPF). The third case study uses a 14-bus system and as security criteria the combined feasibility to the N-1 security constrained AC-OPF and small-signal stability. The fourth case study uses a 162-bus system and as security criteria the feasiblity to the N-1 security constrained AC-OPF under uncertainty. The details of the OPF formulations are provided in the Appendix.} \blue{Please note that in both the formulation of the N-1 security constrained DC- and AC-OPF we do not consider load shedding, as this should be avoided at all times. Here, the goal is to provide the system operator with a tool to rapidly screen a large number of possible operating points and identify possible critical scenarios. For these critical scenarios only, a more dedicated security constrained OPF could be solved to minimize the cost of load shedding or to redispatch generation units.}

To train the neural networks to predict these security criteria, it is necessary to have a dataset of labeled samples that map operating points (neural network inputs $\mb{x}$) to an output security classification $\bar{\mb{y}}$. The neural network predicts the output $\mb{y}$ which should be as close as possible to the ground truth $\bar{\mb{y}}$. We train the neural network as outlined in Section~\ref{sec:NNarchitecture} to achieve satisfactory predictive performance and extract the weights $\mb{W}$ and biases $\mb{b}$ obtained. Then, based on these, we can formulate the verification problems in \eqref{Adv_OPT} and \eqref{Ver_OPT} to derive formal guarantees and assess and improve the robustness of these neural networks including the existence of adversarial examples. For a detailed overview of our methodology please refer to the flowchart in Figure~\ref{flowchart}.

The created dataset can include both historical and simulated operating points. Absent historical data in this paper, we created simulated data for our studies. We will detail the dataset creation for each case study in the corresponding subsection. \blue{To facilitate a more efficient neural network training procedure, we normalize each entry of the input vector $\mb{x}$ to be between 0 and 1 using the upper and lower bounds of the variables (we do that for all samples $\mb{x} \in \mb{S}$). Empirically, this has been shown to improve classifier performance.} 

After creating the training dataset, we export it to TensorFlow \cite{tensorflow2015-whitepaper} for the neural network training. We split the dataset into a training set and a test set. 
We choose to use 85\% of samples for training and 15\% for testing. During training, we minimize the cross-entropy loss function \eqref{cross_entr} using the Adam optimizer with stochastic gradient descent \cite{tensorflow2015-whitepaper}, and use the default options in TensorFlow for training. In the cases where we need to enforce a certain sparsity of the weight matrices we re-train with the same objective function \eqref{cross_entr}, but during training we reduce the number of non-zero weights until a certain level of sparsity is achieved, as explained in Section~\ref{sec:Easier_Ver}. To allow for the neural network verification and the identification of adversarial examples, after the neural network training we export the weight matrices $\mb{W}$ and biases $\mb{b}$ in YALMIP, formulate the MILPs, and solve them with Gurobi. If not noted otherwise, we solve all MILPs to zero optimality gap, and as a result obtain the globally optimal solution. All simulations are carried out on a laptop with processor Intel(R) Core(TM) i7-7820HQ CPU @ 2.90 GHz and 32GB RAM.
  \vspace{-0.3cm}
\subsection{Neural Network Verification for the IEEE 9-bus system} \label{9bus_DCOPF}
\label{sec:9busresults}
\subsubsection{Test Case Setup}
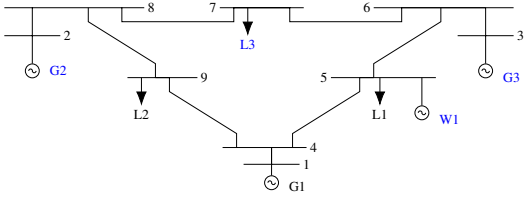
\begin{figure}
\center
\resizebox{7cm}{!}{\begin{tikzpicture}[font = \Large]
\draw
(0,0) node [sin v source] (v1) {} % placing the left most source
(v1.north)--++(0,1) coordinate(v1-up2) % wire going up: store the point
(v1-up2)--++(0,1) coordinate(v1-up) % wire going up: store the point
(v1-up2)--++(-1,0) % left from v1-up
(v1-up2)--++(1,0) node[right]{2} % right from v1-up, setting 1
(v1-up)--++(-1,0) % left from v1-up
(v1-up)--++(4,0) node[right]{8} % right from v1-up, setting 1
coordinate[pos=0.5](v1-r) % identifying first departing point from v1-up
coordinate[pos=0.8](v1-rr) % identifying second departing point from v1-up
% - - - - - - - -
% right part
%(v1-rr)--++(0,-0.25)--++(5,0)--++(0,0.25) % path reaching the right part of the picture
(v1-rr)--++(0,-0.5)--++(4,0) --++(0,0.5) coordinate(v7l)
(v7l) --++ (-0.5,0.0)node[left]{7}
(v7l) --++ (2.5,0.0)
(v7l)--++(2,0)--++(0,-0.5)--++(4,0) --++(0,0.5)
coordinate(v2-ll) % storing the coordinate
(v2-ll)--++(-1,0) node[left]{6} % setting 2
(v2-ll)--++(3,0) coordinate(N6) % going right from our stored coordinate
(N6) --++ (0,-1) coordinate(N3)
(N3) --++ (-1,0) 
(N3) --++ (1,0) node[right]{3}
(N3) --++ (0,-1) node[below,sin v source]{}
(v2-ll)--++(4,0) % going right from our stored coordinate
coordinate[pos=0.5](v2-l) % first departing point right of v2-ll
coordinate[pos=1.5](v2-up) % this will be the up part of the arrow
% - - - - - - - -
% lower part
%(v1-r)--++(0,-0.25)--++(2.625,-2)--++(0,-0.25) % down path (*)
(v1-r)--++(0,-0.5)--++(2.4,-1.5)--++(0,-0.5) coordinate(v9)
(v9) --++ (-1,0)
(v9) --++ (1.5,0) node[right]{9}
(v9) --++ (0.5,0) coordinate(v9r)
(v9r) --++(0,-0.5)--++(2.4,-1.5)--++(0,-0.5)
 coordinate(v3-l) % storing the point
(v3-l)--++(-0.5,0) % going left
(v3-l)--++(2.5,0) node[right]{4} % going right
coordinate[pos=0.5](v3-up) % this will be the up part of the second source 
coordinate[pos=0.8](v3-r) % identifying the point to connect with the right part of the picture
(v3-up) --++(0,-0.6) coordinate(N4)
(N4) --++ (1,0) node[right]{1}
(N4) --++ (-1,0) 
(v3-up) --++(0,-1) node[below,sin v source]{} % attaching the v source
(v3-r) --++(0,0.5)--++(2.4,1.5)--++(0,0.5) coordinate(v5)
(v5) --++ (0.5,0) coordinate(v5r)
(v5r)--++(0,0.5)--++(2.4,1.5)--++(0,0.5) % path to the upper right part
% magically we don' need to replicate in the reverse way (*) completely since
% we have store the coordinate; since we have been smart with the coordinates it's completely simmetric ;) 
;
\path (N3) --++ (0.5,-1.5) node[right]{\textcolor{blue}{G3}};
\path (v1) --++ (0.5,0) node[right]{\textcolor{blue}{G2}};
\path (v3-up) --++ (0.5,-1.4) node[right]{G1};
\path (v3-r) --++(0,0.5)--++(2.622,2) coordinate(N5);
\draw (N5) --++ (2.5,0.0);
\path (N5) --++ (2,0.0) coordinate(W1);
\draw (W1) --++ (0.0,-1) node[below,sin v source]{};
\path (W1) --++ (0.51,-1.5) node[right]{\textcolor{blue}{W1}};
\path (W1) --++ (-1.5,0.0)  coordinate(L1);
\draw[-{Latex[length=0.5cm]},thick](L1)--++(0,-1)node[below]{L1}; % down arrow
\path (v9) --++ (-0.5,0) coordinate(v9lll);
\draw[-{Latex[length=0.5cm]},thick](v9lll)--++(0,-1)node[below]{L2}; % down arrow
\draw (N5) --++ (-1.25,0.0) node[left]{5};
\path (v7l) --++ (0.5,0.0)coordinate(v7m);
\draw[-{Latex[length=0.5cm]},thick](v7m)--++(0,-1)node[below]{\textcolor{blue}{L3}}; % down arrow
%\path (v7l) --++ (1.6,0.0)coordinate(v7mr);
%\draw (v7mr)--++(0,-1)node[below,sin v source]{}; % down arrow
%\path(v7mr)--++(0.5,-1.5)node[right]{{{\color{ETH1}}{W2}}};
\end{tikzpicture}
} 
\vspace{-0.3cm}
\caption{Modified IEEE 9-bus system with wind farm $W1$ at bus 5, and uncertain load L3 at bus 7. Generation and loading marked in blue correspond to the input vector $\mb{x}$ of the neural network.}
\vspace{-0.5cm}
\label{9bus}
\end{figure}
For the first case study, we consider a modified IEEE 9-bus system shown in Fig.~\ref{9bus}. This system has three generators and one wind farm. We assume that the load at bus 7 is uncertain and can vary between 0~MW and 200~MW. The wind farm output is uncertain as well and can vary between 0~MW and 300~MW. The line limits are increased by 25\% to accommodate the added wind generation and increased loading. Remaining parameters are defined according to \cite{zimmerman2010matpower}. \blue{As mentioned in the previous subsection, in this case we employ the N-1 security for the DC power flow as a security criterion.  If the SC-DC-OPF results to infeasibility, then the sample is `unsafe'. We consider the single outage of each of the six transmission lines and use MATPOWER to run the DC power flows for each contingency fixing the generation and load to the defined operating point. We evaluate the violation of active generator and active line limits to determine wheter an operating point is `safe' or `unsafe'.} We create a dataset of 10'000 possible operating points with input vector $\mb{x} =[P_{G2}\,P_{G3}\,P_{L3}\, P_{W1}]$, which are classified $\mb{y} =\{y_1 =\,\emph{safe} \,, \, y_2 = \,\emph{unsafe}\}$. Note that the first generator $P_{G1}$ is the slack bus and its power output is uniquely determined through the dispatch of the other units; therefore, it is not considered as input to the neural network. The possible operating points are sampled using Latin hypercube sampling \cite{mckay1979comparison} which ensures that the distance between each sample is maximized.  

 \bl{To compute the ground-truth (and thus be able to assess the neural network performance) we use a custom implementation of the N-1 preventive security constrained DC-OPF (SC-DC-OPF) in YALMIP using pre-built functions of MATPOWER to compute the bus and line admittance matrix for the intact and outaged system state. We model the the uncertain loads and generators as optimization variables. With ground truth, we refer to the region around an infeasible sample for which we can guarantee that no input exists which is feasible to the SC-DC-OPF problem. We can compute this by minimizing the distance from an infeasible sample to a feasible operation point. For the detailed mathematical formulation of the N-1 security constrained DC-OPF and the ground truth evaluation please refer to Appendix~\ref{App_N1_DCOPF}.} 

\subsubsection{Neural Network Training} Using the created dataset, we train two neural networks which only differ with respect to the enforced sparsity of the weight matrices. Employing loss function \eqref{cross_entr}, the first neural network is trained with dense weight matrices. Using this as a starting point for the retraining, on the second neural network we enforce 80\% sparsity, i.e. only 20\% of the entries are allowed to be non-zero. In both cases, the neural network architecture comprises three hidden layers with 50 neurons each, as this allows us to achieve high accuracy without over-fitting. For comparison, if we would only train a single-layer neural network with 150 neurons, the maximum test set classification accuracy is only 90.7\%, highlighting the need for a multi-layer architecture.

\begin{table}[]
    \caption{Confusion matrices for IEEE 9-bus test case}
    \vspace{-0.1cm}
    \centering
    \begin{tabular}{l | r r | c}
    \toprule
    \multicolumn{4}{c}{Neural network \underline{without} sparsified weight matrices} \\
    \midrule
         Test samples: 1500 & Predicted: Safe & Unsafe & Accuracy\\
         \midrule
         True: Safe (326) & 311 & 15 & \\%  95.4\% \\
         True: Unsafe (1174) & 11  & 1163  \\% & 99.1\%\\
         \midrule
                  &  & & \textbf{98.3\%} \\
        % Accuracy & 97.6\% & 98.7\% & \textbf{98.3\%} \\
         \midrule    \multicolumn{4}{c}{Neural network \underline{with} 80\% sparsified weight matrices} \\
    \midrule
         Test samples: 1500 & Predicted: Safe & Unsafe & Accuracy\\
         \midrule
         True: Safe (326) & 308 & 18 & \\%  94.5\% \\
         True: Unsafe (1174) & 15  & 1159  \\% & 98.7\%\\
         \midrule
       %  Accuracy & 95.4\% & 98.5\% & \textbf{97.8\%} \\
         &  & & \textbf{97.8\%} \\
         \bottomrule
    \end{tabular}
       \label{9bus_NN_CM1}
       \vspace{-0.3cm}
    %\label{9bus_NN_CM2}
\end{table}
Our three-layer neural network without sparsity has a classification accuracy of 98.3\% on the test set and 98.4\% on the training set. This illustrates that neural networks can predict with high accuracy whether an operating point is safe or unsafe. If we allow only 20\% of the weight matrix entries to be non-zero and re-train the three-layer neural network, the classification accuracy evaluates to 97.8\% on the test set and 97.3\% on the training set. The corresponding confusion matrices for both neural networks for the test set are shown in Table~\ref{9bus_NN_CM1}. As we see, the sparsification of the neural network leads to a small reduction of 0.5\% in classification accuracy, since less degrees of freedom are available during training. The benefits from sparsification are, however, substantially more significant. As we will see later in our analysis, sparsification substantially increases the computational tractability of verification problems, and leads to an increase in interpretability of the neural network. 

\subsubsection{Visualization of Verification Samples}
\begin{figure}
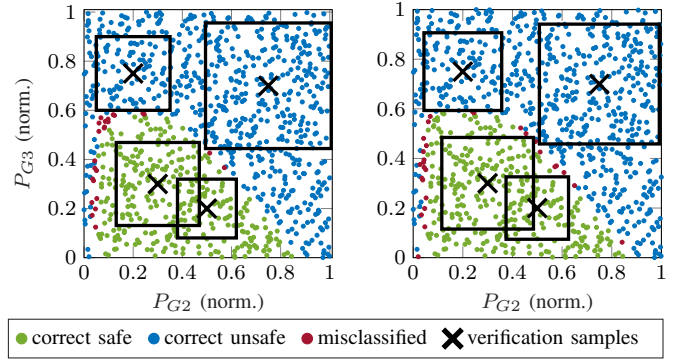

    \begin{footnotesize}
	\input{./Full_Ver.tex} 
			\input{./Spare_Ver.tex} 
%\definecolor{mycolor1}{rgb}{0.00000,0.44700,0.74100}%
%\definecolor{mycolor2}{rgb}{0.85000,0.32500,0.09800}%
         \begin{tikzpicture} 
    \begin{axis}[%
    hide axis,
    xmin=1,
    xmax=2,
    ymin=0,
    ymax=0.1,
    legend style={draw=white!15!black,legend cell align=left},
    legend columns=4
    ]
    \addlegendimage{mark options={solid, mycolor2}, mark=*,only marks}
    \addlegendentry{correct safe \,};
        \addlegendimage{mark options={solid, mycolor1}, mark=*,only marks}
    \addlegendentry{correct unsafe \,};
        \addlegendimage{mark options={solid, mycolor5}, mark=*,only marks}
    \addlegendentry{misclassified \,};
    \addlegendimage{line width=2.0pt,mark=x, mark options={solid, fill=mycolor7, black},only marks,mark size=5.0pt}
    \addlegendentry{verification samples \,};
    \end{axis}
\end{tikzpicture}
    \end{footnotesize}
    \vspace{-0.3cm}
    \caption{Regions around four verification samples in which the classification is guaranteed not to change. The left figure uses the neural network without sparse weight matrices, and the right figure uses the neural network with imposed 80\% sparsity on the weight matrices. For visualization purposes, the load $P_{L3}$ and wind level $P_{W1}$ are fixed to 40\% and 1'000 new samples are created and classified according to the respective neural network.}
    \label{Verification}
    \vspace{-0.3cm}
\end{figure}
To be able to visualize the input regions for which we can guarantee that the trained neural networks will not change their classification, we shall reduce the input dimensionality. For this reason, we will study here only a single uncertainty realization, where we assume that both the load $P_{L3}$ and wind power output $P_{W1}$ amount to 40\% of their normalized output (note that our results apply to all possible inputs). The resulting input space is two-dimensional and includes generators $P_{G2}$ and $P_{G3}$. For visualization purposes only, we take 1'000 new samples from the reduced two-dimensional space and classify them with the neural networks as safe or unsafe. In addition, we compute their true classification using MATPOWER. The resulting classified input spaces are shown in Fig.~\ref{Verification} with the left figure corresponding to the neural network with full weight matrices, and the right figure to the sparsified weight matrices. We can observe that misclassifications occur at the security boundary. This is to be expected as the sampling for this visualization is significantly more dense than the one used for training. What is important here to observe though, is that even if the neural network on the right contains only 20\% of the original number of non-zero weights, the neural networks have visually comparable performance in terms of classification accuracy. As a next step, we solve the verification problem described in \eqref{Ver_OPT}. In that, for several samples $\mb{x}_{\text{ref}}$ we compute the minimum perturbation $\epsilon$ that changes the classification. We visualize the obtained regions using black boxes around the verification samples in Fig.~\ref{Verification}. The average solving time in Gurobi is 5.5~s for the non-sparsified and 0.3~s for the sparsified neural network. We see that by sparsifying the weights we achieve a 20$\times$ computational speed-up. Solving the MILP to zero optimality gap, we are guaranteed that within the region defined by the black boxes around the verification samples no input exists which can change the classification.

\subsubsection{Provable Guarantees}
In the following, we want to provide several provable guarantees of behaviour for both neural networks while considering the entire input space. For this purpose, we solve verification problems of the form \eqref{Ver_OPT}. Since the neural network input $\mb{x}$ is normalized between $\mathbf{0} \leq \mathbf{x} \leq \mathbf{1}$ based on the bounds of the input variables, we include the corresponding constraint \eqref{admis_x} in the verification problem. Similarly, no input to the neural network would violate the limits of the slack bus generator $P_{G1}$. These inputs could be directly classified as unsafe. \blue{ Even if $P_{G1}$ is not part of the input, its limits can be defined based on the input $\mb{x}$ through the DC power balance, as shown in \eqref{DC_PB}:
\begin{align}
    P_{G1}^{\text{min}}  \leq P_{L1} + P_{L2} + P_{L3} - P_{G2} - P_{G3} - P_{W1} \leq P_{G1}^{\text{max}}, \label{DC_PB}
\end{align}
where $\mb{x}=[P_{G2} \,, P_{G3}\,, P_{L3} \,, P_{W1}]^T$. The DC power balance ensures that the system generation equals the system loading. As a result, the slack bus has to compensate the difference between system loading and generation.} In this section, we show how additional a-priori qualifications of the input such as \eqref{DC_PB}, affect the size of the regions in which we can guarantee a certain classification. 
\setlength\tabcolsep{3pt}
\begin{table}[]
\center
    \caption{Verification of Neural Network Properties for 9-bus system}
    \vspace{-0.1cm}
    \label{Prop_9bus}
    \begin{tabular}{l r r r r}
  %  \toprule
    & \multicolumn{2}{c}{w/o power balance} & \multicolumn{2}{c}{with power balance} \\
    \midrule
         & \multicolumn{1}{c}{$\epsilon$} & \multicolumn{1}{c}{Sol. Time (s)} & \multicolumn{1}{c}{$\epsilon$} & \multicolumn{1}{c}{Sol. Time (s)}\\
         \midrule
         \multicolumn{5}{l}{\underline{Property 1:} $\forall x \in [\mathbf{0},\mathbf{1}]: |x - \mathbf{1}|_{\infty} \leq \epsilon$  \quad $\longrightarrow$ Classification: insecure}  \\
          \midrule
         NN (w/o sparsity) & 50.7\% & 2.7 $\mid$ IA: 1.1 & 54.4\% & 1.5  $\mid$ IA: 1.3\\
         NN (80\% sparsity) & 48.7\% & 0.3 $\mid$ IA: 0.2 & 54.4\% & 0.3 $\mid$ IA: 0.2\\
         SC-DC-OPF & 53.7\% & - & 53.7\% & - \\
         \midrule
            \multicolumn{5}{l}{\underline{Property 2:} $\forall x \in [\mathbf{0},\mathbf{1}]: |x - \mathbf{0}|_{\infty} \leq \epsilon$ \quad $\longrightarrow$  Classification: secure } \\
          \midrule
         NN (w/o sparsity) & 29.2\% & 501.9 $\mid$ IA: 400.7 & 29.3\% & 473.0 $\mid$ IA: 817.9 \\
         NN (80\% sparsity) & 32.7\% & 3.3 $\mid$ IA: \quad1.1 & 32.7\% & 2.1 $\mid$ IA: \quad1.6 \\
         SC-DC-OPF\footnotemark & 31.7\% & - & 31.7\% & -\\
         \midrule
            \multicolumn{5}{l}{\underline{Property 3:} $\exists x \in [\mathbf{0},\mathbf{1}]: | P_{W2} - \mathbf{0}|_{\infty} \leq \epsilon$ \quad $\longrightarrow$ Classification: secure }\\
          \midrule
         NN (w/o sparsity) & 97.4\% & 12.2 $\mid$ IA: 11.4 & 90.3\% & 6.5 $\mid$ IA: 3.7 \\
         NN (80\% sparsity) & 99.1\% & 0.4 $\mid$ IA: \,\,\,0.3 & 89.0\% & 0.4 $\mid$ IA: 0.3 \\
         SC-DC-OPF & 92.5\% & - & 92.5\% & - \\
         %\midrule
         \bottomrule
        \multicolumn{5}{l}{${}^2$ The SC-DC-OPF can only provide infeasbility certificates.} \\
       \multicolumn{5}{l}{\, \, We compute this by re-sampling a very large number of samples.}
    \end{tabular}
    \label{Properties_9bus}
    \vspace{-0.3cm}
\end{table}

In Table~\ref{Properties_9bus}, we compare the obtained bounds to the ground truth provided by the SC-DCOPF and report the computational time for three properties, when we do not include and when we do include the power balance constraint. The second reported computational time uses the tightened ReLU bounds \eqref{IA} with interval arithmetic (IA) (see Section~\ref{subsec:IA} for details on the method). The first property is the size of the largest region around the operating point $\mb{x}_{\text{ref}} = \mb{1}$ ($\mathbf{1}$-vector) with the same guaranteed classification. This operating point corresponds to the maximum generation and loading of the system. For this input, the classification of the neural networks is known to be unsafe. We observe for both neural networks that when the power balance constraint is not included in the verification problem, the input region guaranteed to be classified as unsafe is smaller than the ground truth of 53.7\% (provided by the SC-DC-OPF). By including the power balance in the verification problem, the input region classified as unsafe is significantly enlarged and more closely matches the target of 53.7\%.

For the second property, we consider the region around the operating point where all generating units and the load are at their lower bound, i.e., $\mb{x}_{\text{ref}}=\mb{0}$. This point corresponds to a low loading of the system and is therefore secure (i.e. no line overloadings, etc.). We solve the verification problem minimizing the distance from $\mb{x}_{\text{ref}} = \mathbf{0}$ leading to an insecure classification. In this way, we obtain the maximum region described as hyper-cube around the zero vector in which the classification is guaranteed to be always `safe'. We can observe in this case that the neural network without sparsity slightly underestimates while the neural network with 80\% sparsity slightly overestimates the safe region compared to the ground truth (31.7\%). Sparsification allows a $150\times$--$500\times$ computational speed-up (two orders of magnitude). For this property, including the power balance constraint does not change the obtained bounds.

The third property we analyze is the maximum wind infeed for which the neural network identifies a secure dispatch. To this end, we solve the verification problem by maximizing the wind infeed $P_{W1}$ in the objective function while maintaining a secure classification. The true maximum wind infeed obtained by solving this problem directly with the SC-DC-OPF is 92.5\%, i.e. 277.5~MW can be accommodated. If we do not enforce constraint \eqref{DC_PB}, then we observe that the obtained bounds 97.4\% and 99.12\% are not tight. This happens because the obtained solutions $\mb{x}$ from \eqref{Ver_OPT} keep generation and loading to zero and only maximize wind power; this violates the generator bounds on the slack bus, as it has to absorb all the generated wind power. Enforcing the DC power balance \eqref{DC_PB} in the verification problem allows a more accurate representation of the true admissible inputs, and we obtain significantly improved bounds of 90.3\% and 89.0\%. 

For all three properties, we observe that both interval arithmetic (IA) and weight sparsification improve computational tractability, in most instances achieving the lowest solver times when used combined. From the two, weight sparsification has the most substantial effect as it allows to reduce the number of binaries in the resulting MILP problem.

\subsubsection{\blue{Adversarial robustness and re-training}} \label{subsec:Retraining}
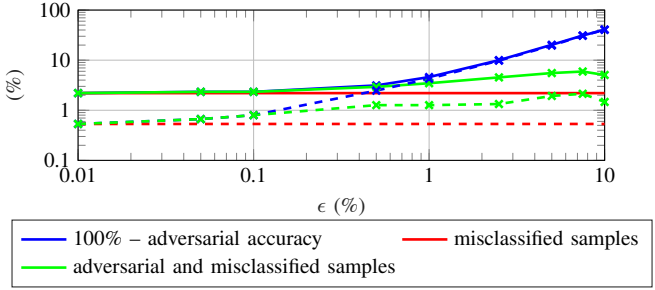
\begin{figure}
    \begin{footnotesize}
\begin{tikzpicture}

\begin{axis}[%
width=7cm,
height=2cm,
scale only axis,
xlabel style={font=\color{white!15!black}},
xlabel={$\epsilon$ (\%)},
xmode=log,
xmin=0.0001,
xmax=0.1,
xminorticks=true,
ymode=log,
ymin=0.1,
ymax=100,
ylabel style={font=\color{white!15!black}},
ylabel={(\%)},
axis background/.style={fill=white},
%legend pos=north west,
xtick={0.0001, 0.001, 0.01, 0.1},
xticklabels={0.01,  0.1, 1, 10},
ytick={0.1,  1, 10, 100},
yticklabels={0.1,  1, 10, 100},
xminorticks=true,
yminorticks=true,
xmajorgrids,
ymajorgrids,
]
\addplot [color=blue, mark=x, mark options={solid, blue}, line width = 1pt]
  table[row sep=crcr]{%
0.0001	2.2\\
0.0005	2.33333333333333\\
0.001	2.33333333333333\\
0.005	3.13333333333333\\
0.01	4.6\\
0.025	10\\
0.05	20.2666666666667\\
0.075	31\\
0.1	40.8\\
};
%\addlegendentry{100\% - Adversarial Accuracy}

\addplot [color=red, line width = 1pt]
  table[row sep=crcr]{%
0.0001	2.2\\
0.0005	2.2\\
0.001	2.2\\
0.005	2.2\\
0.01	2.2\\
0.025	2.2\\
0.05	2.2\\
0.075	2.2\\
0.1	2.2\\
};

%\addlegendentry{Test Set Misclassificiation}

\addplot [color=green, mark=x, mark options={solid, green}, line width = 1pt]
  table[row sep=crcr]{%
0.0001	2.2\\
0.0005	2.33333333333333\\
0.001	2.33333333333333\\
0.005	2.93333333333334\\
0.01	3.46666666666667\\
0.025	4.53333333333333\\
0.05	5.53333333333333\\
0.075	5.93333333333334\\
0.1	5.06666666666666\\
};
%\addlegendentry{Adversarial and Misclassified Examples}
\addplot [color=blue, mark=x, mark options={solid, blue}, line width = 1pt,dashed]
  table[row sep=crcr]{%
0.0001	0.533333333333333\\
0.0005	0.666666666666667\\
0.001	0.8\\
0.005	2.46666666666667\\
0.01	4.26666666666667\\
0.025	9.8\\
0.05	19.6\\
0.075	30.8\\
0.1	40.3333333333333\\
};
%\addlegendentry{100\% - Adversarial Accuracy}

\addplot [color=red, line width = 1pt,dashed]
  table[row sep=crcr]{%
0.0001	0.533333333333333\\
0.0005	0.533333333333333\\
0.001	0.533333333333333\\
0.005	0.533333333333333\\
0.01	0.533333333333333\\
0.025	0.533333333333333\\
0.05	0.533333333333333\\
0.075	0.533333333333333\\
0.1	0.533333333333333\\
};

%\addlegendentry{Test Set Misclassificiation}

\addplot [color=green, mark=x, mark options={solid, green}, line width = 1pt,dashed]
  table[row sep=crcr]{%
0.0001	0.533333333333331\\
0.0005	0.666666666666671\\
0.001	0.799999999999997\\
0.005	1.26666666666667\\
0.01	1.26666666666667\\
0.025	1.33333333333333\\
0.05	1.93333333333334\\
0.075	2.13333333333334\\
0.1	1.46666666666667\\
};
%\addlegendentry{Adversarial and Misclassified Examples}
\end{axis}

\end{tikzpicture}% 
			\vspace{-0.25cm}
			 \begin{center}
			         \begin{tikzpicture}
    \begin{axis}[%
    hide axis,
    xmin=1,
    xmax=2,
    ymin=0,
    ymax=0.1,
    legend style={draw=white!15!black,legend cell align=left},
    legend columns=2
    ]
    \addlegendimage{color=blue, line width = 1pt}
    \addlegendentry{100\% -- adversarial accuracy};
        \addlegendimage{color=red, line width = 1pt}
    \addlegendentry{misclassified samples};
            \addlegendimage{color=green, line width = 1pt}
    \addlegendentry{adversarial and misclassified samples};
    \end{axis}
    \end{tikzpicture}
     \end{center}
    \end{footnotesize}
    \vspace{-0.5cm}
    \caption{\blue{Adversarial accuracy, and share of adversarial and misclassified samples are shown for the test data set of the 9 bus test case and different levels of input perturbation $\epsilon$. The adversarial accuracy refers to the share of correctly classified samples, for which no input perturbation exists which changes the classification within distance $\epsilon$ of that sample (by solving \eqref{Adv_OPT} for each sample). Please note that both axes are logarithmic, and 100\% minus the adversarial accuracy is shown, i.e. the share of samples which are not adversarially robust. Out of these, to determine whether an adversarial example has been identified, the ground-truth classification is computed. The dashed lines show the performance of the re-trained neural network when adding 1'152 identified adversarial examples from the training dataset to the updated training dataset. It can be observed that both prediction accuracy and adversarial robustness, i.e. the share of adversarial examples, are significantly improved.}}
     \vspace{-0.3cm}
    \label{Adversarial_Acc_9bus}
\end{figure}
\blue{In Fig.~\ref{Adversarial_Acc_9bus}, the adversarial robustness and the share of adversarial and misclassified samples are shown for the test dataset and different magnitudes of input perturbation $\epsilon$ in the range from 0.01\% to 10\%, following the methodology described in Fig.~\ref{flowchart}. The full lines are computed using the sparsified neural network with performance described in the confusion matrix in Table~\ref{9bus_NN_CM1}. The sparsified neural network has a predictive accuracy of 97.8\% on the training dataset, i.e. 2.2\% of test samples are misclassified. It can be observed that for small input perturbations $\epsilon$ the share of adversarial and misclassified samples increases to 3.5\%, i.e. for an additional 1.3\% of test samples the identified adversarial input leads to a wrong misclassification. Increasing the input perturbation $\epsilon$ to 10\%, these shares increase to 5.1\% and 2.9\%, respectively. This indicates that parts of the security boundary are not correctly represented by the neural network.} 

\blue{To improve performance, we run the same methodology for the training dataset of 8'500 samples, and include all identified adversarial examples of the training dataset (a total of 1'152 samples) in the updated training dataset and re-train the neural network. Note that we only use the training dataset for this step as the unseen test dataset is used to evaluate the neural network performance. The resulting performance is depicted with dashed lines in Fig.~\ref{Adversarial_Acc_9bus}. Re-training the neural network has two benefits in this test case: First, it improves the predictive accuracy on the unseen test data from 97.8\% to 99.5\%, i.e. only 0.5\% of test samples are misclassified. Second, the share of identified adversarial samples is reduced, for $\epsilon = 1\%$ from 1.3\% to 0.7\% and for $\epsilon = 1\%$ from 2.9\% to 0.9\%, showing improved neural network robustness. At the same time, it can be observed that for perturbations larger than $\epsilon = 1\%$, the adversarial accuracy is similar between both networks. This shows that in this case for a large amount of samples an adversarial input can be identified which correctly leads to a different classification, i.e. the adversarial input moves the operating point across the true security boundary and is not an adversarial example leading to a false misclassification. The neural network robustness could be further improved by repeating this procedure for additional iterations.}

\subsection{Scalability for IEEE 162-bus system} \label{162bus_DCOPF}
\subsubsection{Test Case Setup} For the second case study, we consider the IEEE 162-bus system with parameters taken from \cite{zimmerman2010matpower}. We add five uncertain wind generators located at buses \{3,\,20,\,25,\,80,\,95\} with a rated power of 500~MW each. As security criterion, we consider again N-1 security based on DC power flow, considering the outage of 24 critical lines: $\{22,...,27,144,...,151,272,...,279\}$. The input vector is defined as $\mb{x} = [P_{G_1-G_{12}}\,P_{W_1-W_5}]^T$. To construct the dataset for the neural network training, we first apply a bound tightening of the generator active power bounds, i.e. we maximize and minimize the corresponding bound considering the N-1 \mbox{SC-DC-OPF} constraints. The tightened bounds exclude regions in which the SC-DC-OPF is guaranteed to be infeasible and therefore allows to decrease the input space. In the remaining input space, we draw 10'000 samples using Latin hypercube sampling and classify them as safe or unsafe. As usual in power system problems, the `safe' class is substantially smaller than the `unsafe' class, since the true safe region is only a small subset of the eligible input space. To mitigate the dataset imbalance, we compute the closest feasible dispatch for each of the infeasible samples by solving an SC-DC-OPF problem and using the $\infty$-norm as distance metric. As a result, we obtain a balanced dataset of approximately 20'000 samples.
\subsubsection{Neural Network Training} \begin{table}[]
    \caption{Confusion matrix for IEEE 162-bus test case (with sparsity)}
    \vspace{-0.1cm}
    \centering
    \begin{tabular}{l | r r | c}
    \toprule
         Test samples: 3000 & Predicted: Safe & Unsafe & Accuracy\\
         \midrule
         True: Safe (1507) & 1501 & 6 &  \\
         True: Unsafe (1493) &  11 & 1482 & \\
         \midrule
         Accuracy &  &  & \textbf{99.4}\% \\
         \bottomrule
    \end{tabular}
    \label{162bus_NN_CM1}
    \vspace{-0.1cm}
\end{table}
We choose a neural network architecture with 4 layers of 100 ReLU units at each layer, as the input dimension increases from 4 to 17 compared to the 9-bus system. We train again two neural networks: one without enforcing sparsity and a second with 80\% sparsity. The trained four-layer network without sparsity has a classification accuracy 99.3\% on the test set and 99.7\% on the training set. For the sparsified network, these metrics evaluate to 99.7\% for the test set and 99.4\% for the training set. The confusion matrix of the sparsified network is shown in Table~\ref{162bus_NN_CM1} and it can be observed that the neural network has high accuracy in classifying both safe and unsafe operating points.
\subsubsection{Provable Guarantees} 
\setlength\tabcolsep{3pt}
\begin{table}[]
\center
    \caption{Verification of Neural Network Properties for 162-bus system}
    \vspace{-0.1cm}
    %\label{Prop_162bus}
    \begin{tabular}{l r r r r}
  %  \toprule
    & \multicolumn{2}{c}{w/o power balance} & \multicolumn{2}{c}{with power balance} \\
    \midrule
         & \multicolumn{1}{c}{$\epsilon$} & \multicolumn{1}{c}{Sol. Time (s)} & \multicolumn{1}{c}{$\epsilon$} & \multicolumn{1}{c}{Sol. Time (s)}\\
         \midrule
         \multicolumn{5}{l}{\underline{Property 1:} $\forall x \in [\mathbf{0},\mathbf{1}]: |x - \mathbf{0}|_{\infty} \leq \epsilon$  \quad $\longrightarrow$ Classification: insecure}  \\
          \midrule
    %     NN (w/o sparsity) & 50.7\% & 2.7 & 54.4\% & 1.5\\
      NN (w/o sparsity) & - & $>$ 50 min & - &  $>$ 50 min \\
         NN (80\% sparsity) & 59.4\% & 560 $\mid$ IA: 1217 & 65.5\% & 22 $\mid$ IA: 30\\
         SC-DC-OPF & 66.2\% & - & 66.2\% & - \\
   \bottomrule
    \end{tabular}
    \label{Properties_162bus}
    \vspace{-0.2cm}
\end{table}
Assuming the given load profile, the property of interest is to determine the minimum distance from zero generation $\mb{x}_{\text{ref}} = \mb{0}$ to a feasible (`safe') solution. In Table~\ref{Properties_162bus}, we compare the result of the verification problem with the ground truth obtained from solving the SC-DC-OPF problem. We can see that the bound of 59.4\% we obtain without including the DC power balance is not tight compared to the ground truth of 66.2\%. Including the DC power balance increases the bound to 65.2\% which is reasonably close, indicating satisfactory performance with respect to this property. This confirms the findings for the 9-bus (cmp. Tables~\ref{Prop_9bus}~and~\ref{Properties_162bus}) showing that including the additional power balance constraint in the verification problems leads to bounds on $\epsilon$ closer to the ground-truth. Regarding computational time, we can observe that for this larger neural network, the sparsification of the weight matrices becomes a requirement to achieve tractability. For both cases with and without including the DC power balance, the MILP solver did not identify a solution after 50 minutes for the non-sparsified network. 
\subsubsection{Adversarial Robustness} \begin{figure}
    \begin{footnotesize}
\begin{tikzpicture}

\begin{axis}[%
width=7cm,
height=2cm,
scale only axis,
xlabel style={font=\color{white!15!black}},
xlabel={$\epsilon$ (\%)},
xmode=log,
xmin=0.0001,
xmax=0.1,
xminorticks=true,
ymode=log,
ymin=0.1,
ymax=100,
ylabel style={font=\color{white!15!black}},
ylabel={(\%)},
axis background/.style={fill=white},
%legend pos=north west,
xtick={0.0001, 0.001, 0.01, 0.1},
xticklabels={0.01,  0.1, 1, 10},
ytick={0.1,  1, 10, 100},
yticklabels={0.1,  1, 10, 100},
xminorticks=true,
yminorticks=true,
xmajorgrids,
ymajorgrids,
]
\addplot [color=blue, mark=x, mark options={solid, blue}, line width = 1pt]
  table[row sep=crcr]{%
0.0001	0.6\\
0.0005	0.633333333333333\\
0.001	0.633333333333333\\
0.005	0.833333333333333\\
0.01	0.933333333333333\\
0.025	1.7\\
0.05	5.06666666666667\\
0.075	20.7\\
0.1	44.2\\
};
%\addlegendentry{100\% - Adversarial Accuracy}

\addplot [color=red, line width = 1pt]
  table[row sep=crcr]{%
0.0001	0.566666666666667\\
0.0005	0.566666666666667\\
0.001	0.566666666666667\\
0.005	0.566666666666667\\
0.01	0.566666666666667\\
0.025	0.566666666666667\\
0.05	0.566666666666667\\
0.075	0.566666666666667\\
0.1	0.566666666666667\\
};

%\addlegendentry{Test Set Misclassificiation}

\addplot [color=green, mark=x, mark options={solid, green}, line width = 1pt]
  table[row sep=crcr]{%
0.0001	0.599999999999994\\
0.0005	0.599999999999994\\
0.001	0.599999999999994\\
0.005	0.733333333333334\\
0.01	0.766666666666666\\
0.025	1.16666666666667\\
0.05	2.06666666666666\\
0.075	2.96666666666667\\
0.1	4.86666666666666\\
};
%\addlegendentry{Adversarial and Misclassified Examples}

\end{axis}

\end{tikzpicture}% 
			\vspace{-0.25cm}
			 \begin{center}
			         \begin{tikzpicture}
    \begin{axis}[%
    hide axis,
    xmin=1,
    xmax=2,
    ymin=0,
    ymax=0.1,
    legend style={draw=white!15!black,legend cell align=left},
    legend columns=2
    ]
    \addlegendimage{color=blue, line width = 1pt}
    \addlegendentry{100\% -- adversarial accuracy};
        \addlegendimage{color=red, line width = 1pt}
    \addlegendentry{misclassified samples};
            \addlegendimage{color=green, line width = 1pt}
    \addlegendentry{adversarial and misclassified samples};
    \end{axis}
    \end{tikzpicture}
     \end{center}
    \end{footnotesize}
    \vspace{-0.5cm}
    \caption{\blue{Adversarial accuracy, and share of adversarial and misclassified samples are shown for the test data set of the 162 bus test case and different levels of input perturbation $\epsilon$. The adversarial accuracy refers to the share of correctly classified samples, for which no input perturbation exists which changes the classification within distance $\epsilon$ of that sample (by solving \eqref{Adv_OPT} for each sample). Please note that both axes are logarithmic, and 100\% minus the adversarial accuracy is shown, i.e. the share of samples which are not adversarially robust. Out of these, to determine whether an adversarial example has been identified, the ground-truth classification is computed.}} 
     \vspace{-0.3cm}
    \label{Adversarial_Acc_162busDC}
\end{figure}
\blue{In Fig.~\ref{Adversarial_Acc_162busDC}, the adversarial accuracy, and share of adversarial and misclassified samples are computed for the test data set and different levels of input perturbation $\epsilon$. It can be noted that for small input perturbations $\epsilon\leq1\%$ the adversarial accuracy is above $99\%$ (i.e. for $>99\%$ of test samples no input exists which can change the classification), indicating neural network robustness. For larger input perturbations, the adversarial accuracy decreases as system input perturbations can move the operating point across system security boundaries. The number of identified adversarial examples increases as well. For a large input perturbation $\epsilon=10\%$ for 129 (4.3\%) test samples an adversarial input exists which falsely changes the classification. This indicates that the security boundary estimation of the neural network is inaccurate in some parts of the high-dimensional input space. As for the 9 bus test case, a re-training of the neural network by including identified adversarial examples in the training data set could improve performance.}
\subsection{N-1 Security and Small Signal Stability} \label{14bus_ACOPF}
Security classifiers using neural networks can screen operating points for security assessment several orders of magnitudes faster than conventional methods \cite{arteaga2019deep}. There is, however, a need to obtain guarantees for the behaviour of these classifiers. We show in the following how our framework allows us to analyse the performance of a classifier for both N-1 security and small signal stability.

\subsubsection{Dataset Creation}
 For the third case study, we consider the IEEE 14-bus system \cite{zimmerman2010matpower} and create a dataset of operating points which are classified according to their feasibility to \emph{both} the \mbox{N-1} SC-AC-OPF problem and small signal stability. We consider the outages of all lines except  lines  7-8  and  6-13, as the 14 bus test case is not N-1 secure for these two line outages. Furthermore, we assume that if an outage occurs, the apparent branch flow limits are increased by 12.5\% resembling an emergency overloading capability. We use a simple brute-force sampling strategy to create a dataset of 10'000 equally spaced points in the four input dimensions $\mb{x} = [P_{G2-G5}]$. \blue{The generator automatic voltage regulators (AVRs) are fixed to the set-points defined in \cite{zimmerman2010matpower}. For each of these operating points, we run an AC power flow and the small-signal stability analysis for each contingency. Operating points which satisfy operational constraints and the eigenvalues of the linearized system matrix have negative real parts for all contingencies are classified as 'safe' and 'unsafe' otherwise.} Note that, similar to the 162-bus case, the dataset is unbalanced as only 1.36\% of the overall created samples are feasible. A method to create balanced datasets for \emph{both} static and dynamic security is object of future work. \bl{For the full mathematical details on the N-1 SC-AC-OPF formulation and the small-signal stability constraints, please refer to Appendices~\ref{App_N1_ACOPF} and \ref{Appendix_SSS}, respectively.}

\subsubsection{Neural Network Training \blue{and Performance}} \begin{table}[]
    \caption{Confusion matrix for IEEE 14-bus test case (with sparsity)}
    \vspace{-0.1cm}
    \centering
    \begin{tabular}{l | r r | c}
    \toprule
         Test samples: 1500  & Predicted: Safe & Unsafe & Accuracy\\
         \midrule
         True (15): Safe & 12 & 3 &  \\
         True (1485): Unsafe &  0 & 1485 & \\
         \midrule
         Accuracy &  &  & 99.8\% \\
         \bottomrule
    \end{tabular}
    \label{14bus_NN_CM1}
     \vspace{-0.1cm}
\end{table}
\begin{figure}
    \begin{footnotesize}
\begin{tikzpicture}

\begin{axis}[%
width=7cm,
height=2cm,
scale only axis,
xlabel style={font=\color{white!15!black}},
xlabel={$\epsilon$ (\%)},
xmode=log,
xmin=0.0001,
xmax=0.1,
xminorticks=true,
ymode=log,
ymin=0.1,
ymax=100,
ylabel style={font=\color{white!15!black}},
ylabel={(\%)},
axis background/.style={fill=white},
%legend pos=north west,
xtick={0.0001, 0.001, 0.01, 0.1},
xticklabels={0.01,  0.1, 1, 10},
ytick={0.1,  1, 10, 100},
yticklabels={0.1,  1, 10, 100},
xminorticks=true,
yminorticks=true,
xmajorgrids,
ymajorgrids,
]
\addplot [color=blue, mark=x, mark options={solid, blue}, line width = 1pt]
  table[row sep=crcr]{%
0.0001	0.2\\
0.0005	0.2\\
0.001	0.2\\
0.005	0.266666666666667\\
0.01	0.466666666666667\\
0.025	0.933333333333333\\
0.05	1.93333333333333\\
0.075	3\\
0.1	4.13333333333333\\
};
%\addlegendentry{100\% - Adversarial Accuracy}

\addplot [color=red, line width = 1pt]
  table[row sep=crcr]{%
0.0001	0.2\\
0.0005	0.2\\
0.001	0.2\\
0.005	0.2\\
0.01	0.2\\
0.025	0.2\\
0.05	0.2\\
0.075	0.2\\
0.1	0.2\\
};

%\addlegendentry{Test Set Misclassificiation}

\addplot [color=green, mark=x, mark options={solid, green}, line width = 1pt]
  table[row sep=crcr]{%
0.0001	0.200000000000003\\
0.0005	0.200000000000003\\
0.001	0.200000000000003\\
0.005	0.266666666666666\\
0.01	0.266666666666666\\
0.025	0.400000000000006\\
0.05	0.400000000000006\\
0.075	0.666666666666671\\
0.1	0.933333333333337\\
};
%\addlegendentry{Adversarial and Misclassified Examples}

\end{axis}

\end{tikzpicture}% 
			\vspace{-0.25cm}
			 \begin{center}
			         \begin{tikzpicture}
    \begin{axis}[%
    hide axis,
    xmin=1,
    xmax=2,
    ymin=0,
    ymax=0.1,
    legend style={draw=white!15!black,legend cell align=left},
    legend columns=2
    ]
    \addlegendimage{color=blue, line width = 1pt}
    \addlegendentry{100\% -- adversarial accuracy};
        \addlegendimage{color=red, line width = 1pt}
    \addlegendentry{misclassified samples};
            \addlegendimage{color=green, line width = 1pt}
    \addlegendentry{adversarial and misclassified samples};
    \end{axis}
    \end{tikzpicture}
     \end{center}
    \end{footnotesize}
    \vspace{-0.5cm}
    \caption{\blue{Adversarial accuracy, and share of adversarial and misclassified samples are shown for the test data set of the 14 bus test case and different levels of input perturbation $\epsilon$. The adversarial accuracy refers to the share of correctly classified samples, for which no input perturbation exists which changes the classification within distance $\epsilon$ of that sample (by solving \eqref{Adv_OPT} for each sample). Please note that both axes are logarithmic, and 100\% minus the adversarial accuracy is shown, i.e. the share of samples which are not adversarially robust. Out of these, to determine whether an adversarial example has been identified, the ground-truth classification is computed.}} 
     \vspace{-0.3cm}
    \label{Adversarial_Acc}
\end{figure}
We choose a three-layer neural network with 50 neurons for each layer, as in Section~\ref{sec:9busresults}. Based on the analysis of the previous case studies, here we only train an 80\% sparsified network. The network achieves the same accuracy of 99.8\% both on the training and on the test set. The confusion matrix on the test set is shown in Table~\ref{14bus_NN_CM1}. Note that here the accuracy does not carry sufficient information as a metric of the neural network performance, as simply classifying all samples as unsafe would lead to a test set accuracy of 99.0\% due to the unbalanced classes. Besides the use of supplementary metrics, such as specificity and recall (see e.g. \cite{arteaga2019deep}), obtaining provable guarantees of the neural network behavior becomes, therefore, of high importance. 

\blue{One of the main motivations for data-driven approaches are their computational speed-up compared to conventional methods. To assess the computational complexity, we compare the computational time required for evaluating 1'000 randomly sampled operating points using AC power flows and the small-signal model to using the trained neural network to predict power system security. We find that, on average, computing the AC power flows and small-signal model for all contingencies takes 0.22~s, and evaluating the neural network $7\times10^{-5}$~s, translating to a computational speed up of factor 1000 (three orders of magnitude). Similar computational speed-ups are reported in \cite{arteaga2019deep, du2019achieving} for deep learning applications to system security assessment. Note that the works in \cite{arteaga2019deep, du2019achieving} do not provide performance guarantees and do not examine robustness. Contrary, in the following, we provide formal guarantees for the performance of the security classifier by analysing the adversarial accuracy and identifying adversarial examples.}
\subsubsection{Evaluating Adversarial Accuracy} \label{sec:Adv_Acc_Res}
Adversarial accuracy identifies the number of samples whose classification changes from the ground-truth classification if we perform a perturbation to their input. Assuming that points in the neighborhood of a given sample should share the same classification, e.g. points around a `safe' sample would likely be safe, a possible classification change indicates that we are either very close to the security boundary or we might have discovered an adversarial example. Carrying out this procedure for our whole test dataset, we would expect that in most cases the classification in the vicinity of the sample will not change (except for the comparably small number of samples that is close to the security boundary). In Fig.~\ref{Adversarial_Acc} the adversarial accuracy is depicted for the sparsified neural network. It can be observed that for small perturbations, i.e. $\epsilon \leq 1\%$, the adversarial accuracy stays well above 99\%; that is, for 99\% of test samples no input exists within distance of $\epsilon$ which changes the classification. Only if large perturbations to the input are allowed (i.e. $\epsilon \geq 1\%$) the adversarial accuracy decreases. This shows that the classification of our neural network is adversarially robust in most instances. Note that the adversarial accuracy only makes a statement regarding the classification by the neural network and is not related to the ground truth classification. In a subsequent step, as we show in the next paragraph, we need to assess whether identified samples are in fact misclassified.
\subsubsection{Identifying Adversarial Examples}
\begin{figure}
    \begin{footnotesize}
			\input{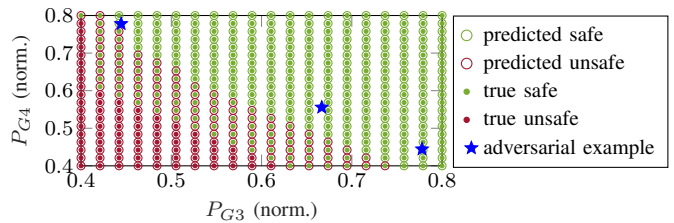} 
    \end{footnotesize}
    \vspace{-0.75cm}
    \caption{Classification of 400 new equally spaced samples for the IEEE 14-bus system. For 2-D visualization purposes, the active power of $P_{G2}$ and $P_{G5}$ are fixed to their maximum output. \blue{Blue stars mark identified adversarial examples. For these samples a small input perturbation exist which falsely changes the classification.} The reason is the inaccurate prediction of the system security boundary.} %The original 100 data set samples
    \label{Adv_example}
   \vspace{-0.4cm}
\end{figure}
Having identified regions where adversarial examples may exist, we can now proceed with determining if they truly exist. \blue{The resulting share of identified adversarial examples is shown in Fig.~\ref{Adversarial_Acc}. Focusing on the test samples which are not adversarially robust for small $\epsilon$, i.e. $\epsilon \leq 1\%$, we identify an adversarial example for the sample $\mb{x} = [1.0\,1.0\,0.4444\,0.7778]$ with classification `safe'. Modifying this input only by $\epsilon=0.5\%$, we identify the adversarial input $\mb{x}_{\text{adv}} = [0.9950\,0.9950\,0.4394\,0.7728]$ which falsely changes the classification to `unsafe', i.e. $y_{\text{adv},1} > y_{\text{adv},2}$. Allowing an input modification of magnitude $\epsilon = 1\%$, we identify two additional adversarial examples for inputs $\mb{x} = [1.0\,1.0\,0.6667\,0.5556]$ and $\mb{x} = [1.0\,1.0\,0.7778\,0.4444]$ with classification `safe', respectively. These have the corresponding adversarial inputs  $\mb{x}_{\text{adv}} = [0.9750\,0.9750\,0.6417\,0.5306]$ and  $\mb{x}_{\text{adv}} = [0.9750\,0.9750\,0.7528\,0.4194]$ which falsely change the classification to `unsafe', respectively.} 

For illustrative purposes in Fig.~\ref{Adv_example}, we re-sample 400 equally spaced samples and compute both the neural network prediction and ground truth. In Fig.~\ref{Adv_example} the location of the adversarial examples is marked by a star. We can observe that the neural network boundary prediction is not precise and as a result, the adversarial inputs get falsely classified as unsafe. This highlights the additional benefit of the presented framework to identify adversarial examples, and subsequently regions in which additional detailed sampling for re-training the classifier is necessary. 
\vspace{-0.3cm}

\subsection{\blue{N-1 Security and Uncertainty}}  \label{162bus_ACOPF}
\blue{For the fourth case study, to further demonstrate scalability of our methodology, we use the IEEE 162 bus test case with parameters defined in \cite{PGLIB}, and train a neural network to predict power system security with respect to the N-1 security constrained AC-OPF under uncertainty. Compared to the previous 14 bus test case, we assume that the voltage set-points of generators can vary within their defined limits (i.e. they are part of the input vector $\mb{x}$), and we consider both uncertain injections in power generation and demand. For the N-1 security assessment, we consider the possible outages of lines $\{6, 8, 24, 50, 128\}$, assuming the same parameters for the outaged system state as for the intact system state. Furthermore, we place 3 wind farms with rated power of 500 MW and consider 3 uncertain loads with $\pm50$\% variability, i.e., a total of 6 uncertain power injections, at buses $\{60, 90, 145, 3, 8, 52\}$. For all uncertain injections, we assume a power factor $\cos \phi=1$.}
\subsubsection{\blue{Dataset creation}} \blue{As the resulting input dimension of the considered test case is high (29 inputs) and large parts of the input space correspond to infeasible operating points, a sampling strategy based on Latin hypercube sampling from the entire input space is not successful at recovering feasible samples. To create the dataset, we rely on an efficient dataset creation method proposed in \cite{venzke2019efficient}. The dataset creation method consists of several steps. First, all the upper and lower bounds in the AC-OPF problem and on the input variables in $\mb{x}$ are tightened using bound tightening algorithms in \cite{Shchetinin2019} and \cite{sundar2018optimization}. 
Second, relying on infeasibility certificates based on convex relaxations of the AC-OPF, large parts of the input space can be characterized as infeasible, i.e. `unsafe' with respect to the power system security criteria. These infeasibility certificates take the form of hyperplanes and the remaining unclassified space can be described as a convex polyhedron $\mb{A} \mb{x} \leq \mb{b}$. Third, a large number of samples (here 10'000) are sampled uniformly from inside the convex polyhedron and their classification is computed. For infeasible samples, the closest feasible sample is determined to characterize the security boundary. Finally, a Gaussian distribution is fitted to the feasible boundary samples, and an additional large number of samples (here 100'000) are sampled from this distribution and their feasibility to the N-1 security constrained AC-OPF is assessed. This methodology results to a database of 121'358 samples out of which 23.2\% correspond to `safe' operating points. Note that computing the infeasibility certificates allows to reduce the considered input volume from a normalized volume of 1 (i.e. all bounds on $\mb{x}$ are between 0 and 1) to a volume of $6 \times 10^{-10}$. This highlights the need for advanced methods for dataset creation, as direct sampling strategies are not able to produce balanced datasets balanced of `safe' and `unsafe' operating points. For more details on the dataset creation method, for brevity, the reader is referred to \cite{venzke2019efficient}.}
\subsubsection{\blue{Neural Network Training and Adversarial Robustness}} 
\begin{table}[]
 \caption{Confusion matrix for IEEE 162-bus test case (N-1 security AC-OPF and Uncertainty)}
    \vspace{-0.1cm}
    \centering
    \begin{tabular}{l | r r | c}
    \toprule
         Test samples: 18204 & Predicted: Safe & Unsafe & Accuracy\\
         \midrule
         True: Safe (4260) & 3148 & 1112 &  \\
         True: Unsafe (13944) &  609 & 13335 & \\
         \midrule
         Accuracy &  &  & \textbf{90.5}\% \\
         \bottomrule
    \end{tabular}
    \label{162bus_AC_CM}
    \vspace{-0.1cm}
\end{table}

\begin{figure}
    \begin{footnotesize}
\begin{tikzpicture}

\begin{axis}[%
width=7cm,
height=2cm,
scale only axis,
xlabel style={font=\color{white!15!black}},
xlabel={$\epsilon$ (\%)},
xmode=log,
xmin=0.0001,
xmax=0.1,
xminorticks=true,
ymode=log,
ymin=5,
ymax=100,
ylabel style={font=\color{white!15!black}},
ylabel={(\%)},
axis background/.style={fill=white},
%legend pos=north west,
xtick={0.0001, 0.001, 0.01, 0.1},
xticklabels={0.01,  0.1, 1, 10},
ytick={10, 100},
yticklabels={10, 100},
xminorticks=true,
yminorticks=true,
xmajorgrids=true,
ymajorgrids=true,
]
\addplot [color=blue, mark=x, mark options={solid, blue}, line width = 1pt]
  table[row sep=crcr]{%
0.0001	10.6460118655241\\
0.0005	16.1557899362777\\
0.001	24.6868820039552\\
0.005	82.5697648868381\\
0.01	89.991210722918\\
0.025	91.5128543177324\\
0.05	91.6941331575478\\
0.1     94.04 \\
};
%\addlegendentry{100\% - Adversarial Accuracy}

\addplot [color=red, line width = 1pt]
  table[row sep=crcr]{%
0.0001	9.45396616128324\\
0.0005	9.45396616128324\\
0.001	9.45396616128324\\
0.005	9.45396616128324\\
0.01	9.45396616128324\\
0.025	9.45396616128324\\
0.05	9.45396616128324\\
0.1     9.45396616128324\\
};

%\addlegendentry{Test Set Misclassificiation}

\addplot [color=green, mark=x, mark options={solid, green}, line width = 1pt]
  table[row sep=crcr]{%
0.0001	10.5251593056471\\
0.0005	13.4585805317513\\
0.001	14.6835860250494\\
0.005	28.125686662272\\
0.01	42.1500769061745\\
0.025	73.9782465392221\\
0.05	74.4012304987915\\
0.1     76.75\\
};
%\addlegendentry{Adversarial and Misclassified Examples}

\end{axis}

\end{tikzpicture}% 
			\vspace{-0.25cm}
			 \begin{center}
			         \begin{tikzpicture}
    \begin{axis}[%
    hide axis,
    xmin=1,
    xmax=2,
    ymin=0,
    ymax=0.1,
    legend style={draw=white!15!black,legend cell align=left},
    legend columns=2
    ]
    \addlegendimage{color=blue, line width = 1pt}
    \addlegendentry{100\% -- adversarial accuracy};
        \addlegendimage{color=red, line width = 1pt}
    \addlegendentry{misclassified samples};
            \addlegendimage{color=green, line width = 1pt}
    \addlegendentry{adversarial and misclassified samples};
    \end{axis}
    \end{tikzpicture}
     \end{center}
    \end{footnotesize}
    \vspace{-0.5cm}
    \caption{\blue{Adversarial accuracy, and share of adversarial and misclassified samples are shown for the test data set of the 162 bus test case (N-1 security and uncertainty) and different levels of input perturbation $\epsilon$. The adversarial accuracy refers to the share of correctly classified samples, for which no input perturbation exists which changes the classification within distance $\epsilon$ of that sample (by solving \eqref{Adv_OPT} for each sample). Please note that both axes are logarithmic, and 100\% minus the adversarial accuracy is shown, i.e. the share of samples which are not adversarially robust. Out of these, to determine whether an adversarial example has been identified, the ground-truth classification is computed.}}
     \vspace{-0.3cm}
    \label{Adversarial_Acc_162busAC}
\end{figure}
\blue{We train a neural network with 3 hidden layers of 100 neurons each and enforce a weight sparsity of 80\%. The neural network has an accuracy of 91.3\% on the training dataset and 90.5\% on the test dataset. The confusion matrix for the test dataset is shown in Table~\ref{162bus_AC_CM}.} 
\blue{Similar to previous test cases, we evaluate the adversarial accuracy in Fig.~\ref{Adversarial_Acc_162busAC}. 
We find that the neural network is not adversarially robust, already for an input modification of $\epsilon=0.1\%$ for 4.2\% of test samples an adversarial example is identified (in addition to the initially misclassified 9.5\% of test samples). For an input modification of $\epsilon=1\%$ this value increases to 31.6\%. This systematic process to identify adversarial examples proposed in this paper allows us to obtain additional insights about the quality of the training database. Assessing the adversarial accuracy (i.e. the number of samples that change classification within a distance $\epsilon$) versus the actual adversarial examples, we find that the change in classification in this case often occurs because the operating points have been moved across the true system security boundary. This indicates that many samples are placed close to the correctly predicted true system boundary in the high-dimensional space. Using high-performance computing, an additional detailed re-sampling of the system security boundary or re-training by including adversarial examples as shown in Section~\ref{subsec:Retraining} could improve neural network robustness.} 
\section{Conclusion}
Neural networks in power system applications have so far been treated as a black box; this has become a major barrier towards their application in practice. This is the first work to present a rigorous framework for neural network verification in power systems and to obtain provable performance guarantees. To this end, we formulate verification problems as mixed-integer linear programs and train neural networks to be computationally easier verifiable. We provably determine the range of inputs that are classified as safe or unsafe, and \blue{systematically} identify adversarial examples, i.e. slight modifications in the input that lead to a mis-classification by neural networks. This enables power system operators to understand and anticipate the neural network behavior, building trust in them, and remove a major barrier toward their adoption in power systems. We verify properties of a security classifier for an IEEE \mbox{9-bus} system, \blue{improve its robustness} and demonstrate its scalability for a \mbox{162-bus} system, highlighting the need for sparsification of neural network weights. Finally, we further identify adversarial examples and evaluate the adversarial accuracy of neural networks trained to assess N-1 security \blue{under uncertainty} and small-signal stability.

 \appendices

\section{Optimal Power Flow (OPF) Formulations}
\bl{In the following, for completeness, we provide a brief overview of the DC and AC optimal power flow formulations including N-1 security and small-signal stability. For more details please refer to \cite{cain2012history, stott2009dc,capitanescu2011state}.}

\subsection{Preliminaries} \bl{We consider a power grid which consists of buses (denoted with the set $\mathcal{N}$) and transmission lines (denoted with the set $\mathcal{L}$). The transmission lines connect one bus $i \in \mathcal{N}$ to another bus $j \in \mathcal{N}$, i.e., $(i,j) \in \mathcal{L}$. For the AC-OPF formulation, we consider the following variables for each bus: The voltage magnitudes $\bm{V}$, the voltage angles $\bm{\theta}$, the active power generation $\bm{P_G}$, the reactive power generation $\bm{Q_G}$, the active wind power generation $\bm{P_W}$, and the active and reactive power demand $\bm{P_L}$ and $\bm{Q_L}$. Each of these vectors have the size $n_b \times 1$, where $n_b$ is the number of buses in the set $\mathcal{N}$. Note that during operation, for one specific instance, the active wind power generation and active and reactive power demand are assumed to be fixed (and the curtailment of wind generators and load shedding are to be avoided at all times). To comply with the N-1 security criterion, we consider the possible single outage of a set of lines, which we denote with the set $\mathcal{C}$. The first element of this set corresponds to the intact system state, denoted with '0'. The superscript 'c' denotes the variables corresponding to the intact and outaged system states.}

\subsection{N-1 Security-Constrained DC-OPF} \label{App_N1_DCOPF}
\bl{In Section~\ref{9bus_DCOPF} and Section~\ref{162bus_DCOPF}, we create datasets of operating points classified according to their feasibility to the N-1 security constrained DC-OPF. The DC-OPF approximation neglects reactive power and active power losses, and assumes that the voltage magnitudes $\bm{V}$ are fixed, for a detailed treatment please refer to \cite{stott2009dc}. Considering a set  $\mathcal{C}$ of possible single line outages, the preventive N-1 security constrained DC-OPF problem can then be formulated as follows:
\begin{alignat}{3}
    \min_{\bm{P_G^c}, \bm{P_L}, \bm{P_W}, \bm{\theta}^c}\, \, & \bm{f}(\bm{P_G}^0) && \\
   \text{s.t.} \, \, & \bm{P_G^c} + \bm{P_W} - \bm{P_L} = \bm{B_{\text{bus}}^c} \bm{\theta^c} && \quad \forall c \in \mathcal{C} \label{nodal} \\
   & \bm{P_{\text{line}}^{\text{min},c}} \leq \bm{B_{\text{line}}^c} \bm{\theta^c} \leq \bm{P_{\text{line}}^{\text{max},c}} && \quad \forall c \in \mathcal{C}  \label{Pline_lim} \\
      & \bm{P_{G}^{\text{min}}} \leq \bm{P_G^c} \leq \bm{P_{G}^{\text{max}}} && \quad  \forall c \in \mathcal{C} \label{PG_lim}\\
      & \bm{P_{W}^{\text{min}}} \leq \bm{P_W} \leq \bm{P_{W}^{\text{max}}} &&  \label{PW_lim}\\
      & \bm{P_{L}^{\text{min}}} \leq \bm{P_L} \leq \bm{P_{L}^{\text{max}}}  &&  \label{PL_lim}\\
      & \bm{P_G}^{\text{wosb},0} = \bm{P_G}^{\text{wosb},c} && \quad  \forall c \in \mathcal{C} \label{PG_prev}
\end{alignat}
The objective function $\bm{f}$ minimizes e.g. the generation cost of the intact system state. The nodal power balance in \eqref{nodal} has to be satisfied for the intact and outaged system states. The bus and line admittance matrices are denoted with $\bm{B_{\text{bus}}}$ and $\bm{B_{\text{line}}}$, respectively. Upper and lower limits are enforced for the active power line flows, generation, wind power, and load demands in \eqref{Pline_lim}, \eqref{PG_lim}, \eqref{PW_lim} and \eqref{PL_lim}, respectively. The constraint in \eqref{PG_prev} enforces preventive control action of generators. The superscript 'wosb' denotes all generators except the slack bus generator. The independent variables characterizing an operating point are $\mb{x}:=[\bm{P_G}^{\text{wosb},0}, \bm{P_W}, \bm{P_L}]$. To create datasets, we compute the classification for each operation point by first running DC power flows to determine $\bm{\theta^c}$ and the slack bus generator dispatch $\bm{P_G}^{\text{slack},c}$ for the intact and each outaged system states. The superscript `slack' denotes the slack bus. Then, we check satisfaction of the constraints on active generator power \eqref{PG_lim} and active line flows \eqref{Pline_lim}. In operations, it is usually assumed that both the wind power and loading are fixed, i.e. $\bm{P_{W}^{\text{max}}}=\bm{P_{W}^{\text{min}}}$ and $\bm{P_{L}^{\text{max}}}=\bm{P_{L}^{\text{min}}}$. Here, we model them as variables to be able to compute the ground-truth for the region around an infeasible sample $\mb{x}^{\text{infeas}}$ in which no feasible sample exist. To this end, we solve the following optimization problem computing the minimum distance from the infeasible sample to an operating point that is feasible to the N-1 security-constrained DC-OPF:
\begin{align}
        \min_{\bm{P_G^c}, \bm{P_L}, \bm{P_W}, \bm{\theta}^c}\, \, & |\mb{x} - \mb{x}^{\text{infeas}}|_\infty \\
        \text{s.t.} \, \, & \eqref{nodal} - \eqref{PG_prev} \\
        & \mb{x}=[\bm{P_G}^{\text{wosb},0}, \bm{P_W}, \bm{P_L}] 
\end{align}
As this optimization problem is convex, we can provably identify the closest feasible sample $\mb{x}$ to $\mb{x}^{\text{infeas}}$. Note that we solve this optimization problem to compute the results denoted with 'SC-DC-OPF' in Table~\ref{Properties_9bus} and Table~\ref{Properties_162bus}.}
\subsection{N-1 Security-Constrained AC-OPF} \label{App_N1_ACOPF}
\bl{In Section~\ref{14bus_ACOPF} and Section~\ref{162bus_ACOPF}, we create datasets of operating points classified according to their feasibility to the N-1 security constrained AC-OPF. In Section~\ref{14bus_ACOPF}, we consider small-signal stability constraints as well. These will be discussed in the Appendix~\ref{Appendix_SSS}. We can formulate the N-1 security constrained AC-OPF problem as follows:
\begin{alignat}{3}
\min_{\mb{z}^c} \, \, & f(\mb{P_g^0}) && \label{objac}\\
    \text{s.t.}  \quad           & \mb{z}^c := \{ \bm{V^c},\bm{\theta^c},\bm{P_G^c},\bm{Q_G^c},\bm{P_W},\bm{P_L},\bm{Q_L} \} && \quad  \forall c \in \mathcal{C} \label{z_def} \\
    & \mb{s}_{\text{balance}} (\mb{z}^c) = \mb{0} && \quad \forall c \in \mathcal{C} \label{con_acpf} \\
        & | \mb{s}_{\text{line}} (\mb{z}^c) | \leq \mb{s}_{\text{line}}^{\text{max},c} && \quad \forall c \in \mathcal{C} \label{con_s} \\
    &  \bm{P_{G}^{\text{min}}} \leq \bm{P_G^c} \leq \bm{P_{G}^{\text{max}}} && \quad  \forall c \in \mathcal{C} \label{PG_limac}\\
        &  \bm{Q_{G}^{\text{min}}} \leq \bm{Q_G^c} \leq \bm{Q_{G}^{\text{max}}} && \quad  \forall c \in \mathcal{C} \label{QG_limac}\\
      & \bm{P_{W}^{\text{min}}} \leq \bm{P_W} \leq \bm{P_{W}^{\text{max}}} && \quad  \label{PW_limac}\\
      & \bm{P_{L}^{\text{min}}} \leq \bm{P_L} \leq \bm{P_{L}^{\text{max}}}  && \quad   \label{PL_limac}\\
            & \bm{Q_{L}^{\text{min}}} \leq \bm{Q_L} \leq \bm{Q_{L}^{\text{max}}}  && \quad   \label{QL_limac}\\
    & \mb{V}^{\text{min}} \leq \mb{V}^c \leq \mb{V}^{\text{max}} && \quad \forall c \in \mathcal{C} \label{con_v} \\
         & \bm{P_G}^{\text{wosb},0} = \bm{P_G}^{\text{wosb},c}, \bm{V_G}^c = \bm{V_G}^c && \quad  \forall c \in \mathcal{C} \label{link} 
\end{alignat}
The vector $\mb{z}^c$ collects all variables for the intact and outaged system states in \eqref{z_def}. The non-linear AC power flow nodal balance $\mb{s}_{\text{balance}}$ in \eqref{con_acpf} has to hold for intact and outaged system states. The absolute apparent line flow $|\mb{s}_{\text{line}}|$ is constrained by an upper limit in \eqref{con_s}. For the full mathematical formulation, for brevity, please refer to \cite{cain2012history}. The upper and lower limits on the system variables are defined in the constraints \eqref{PG_limac} -- \eqref{con_v}. The constraint \eqref{link} enforces preventive control actions for N-1 security: Both the generator active power set-points and voltage set-points remain fixed during an outage. Note that the vector $\bm{V_G}$ refers to the voltage set-points of the generators. We do not fix the entry in $\bm{P_G}$ corresponding to the slack bus, as the slack bus generator compensates the mismatch in the losses. The independent variables that characterize an operating point are $\mb{x}: =[\bm{P_G}^{\text{wosb},0}, \bm{P_W}, \bm{P_L},\bm{Q_L},\bm{V_G}^0]$. To create the datasets, based on the operating point $\mb{x}$, we solve the AC power flow equations in \eqref{con_acpf} to determine the dependent variables for each contingency and the intact system state. Then, we check the satisfaction of the operational constraints of the \mbox{N-1} SC-AC-OPF problem including active and reactive generator limits \eqref{PG_limac} and \eqref{QG_limac}, apparent branch flow limits \eqref{con_s}, and voltage limits \eqref{con_v}. For an operating point to be classified as feasible to the the \mbox{N-1} SC-AC-OPF problem, it must satisfy all these constraints for all considered contingencies.} % This can be achieved by running an AC power flow for each contingency with the variables fixed to correspond to the defined operating point, and checking constraint violations. 

\subsection{Small-Signal Stability} \label{Appendix_SSS}
\bl{For the IEEE 14 bus test case in Section~\ref{14bus_ACOPF} we evaluate the feasibility of operating points with respect to combined small-signal stability and N-1 security. For the small signal stability model, we rely on standard system models and parameters commonly used for small signal stability analysis defined in Refs.~\cite{sauer1998power} and \cite{milano2010power}. We use a sixth-order synchronous machine model with an Automatic Voltage Regulator (AVR) Type I with three states. A more detailed description of the system model and parameters can be found in the Appendix of Ref.~\cite{thams2017data}. To determine small-signal stability, we linearize the dynamic model of the system around the current operating point, and compute the eigenvalues $\bm{\lambda}$ of the resulting system matrix $\mb{A}$. If all eigenvalues have negative real parts (i.e. lie in the left-hand plane), the system is considered small-signal stable, otherwise unstable. This can be formalized as follows:
\begin{alignat}{3}
      & \mb{A}(\mb{z^c}) \bm{\nu}^c = \bm{\lambda}^c \bm{\nu}^c && \quad \forall c \in \mathcal{C}  \label{SS1} \\
      & \bm{\lambda}^c \leq \mb{0} &&  \quad \forall c \in \mathcal{C}   \label{SS2} 
\end{alignat}
The set of variables $\mb{z^c}$ is defined in \eqref{z_def}. The term $\bm{\nu}^c$ denotes the right-hand eigenvectors of the system matrix $\mb{A}$. As we consider both N-1 security and small-signal stability, we have to modify the small-signal stability model for each operating point and contingency. We use Mathematica to derive the small signal model symbolically, MATPOWER AC power flows to initialize the system matrix, and Matlab to compute its eigenvalues and damping ratio, and assess the small-signal stability for each operating point and contingency.}

\bibliographystyle{IEEEtran}

\bibliography{Bib}

\newpage 
%\vspace{-7cm}
\ifCLASSOPTIONcaptionsoff
\newpage
\fi
\begin{IEEEbiography}[{\includegraphics[width=1in,height=1.25in,clip,keepaspectratio]{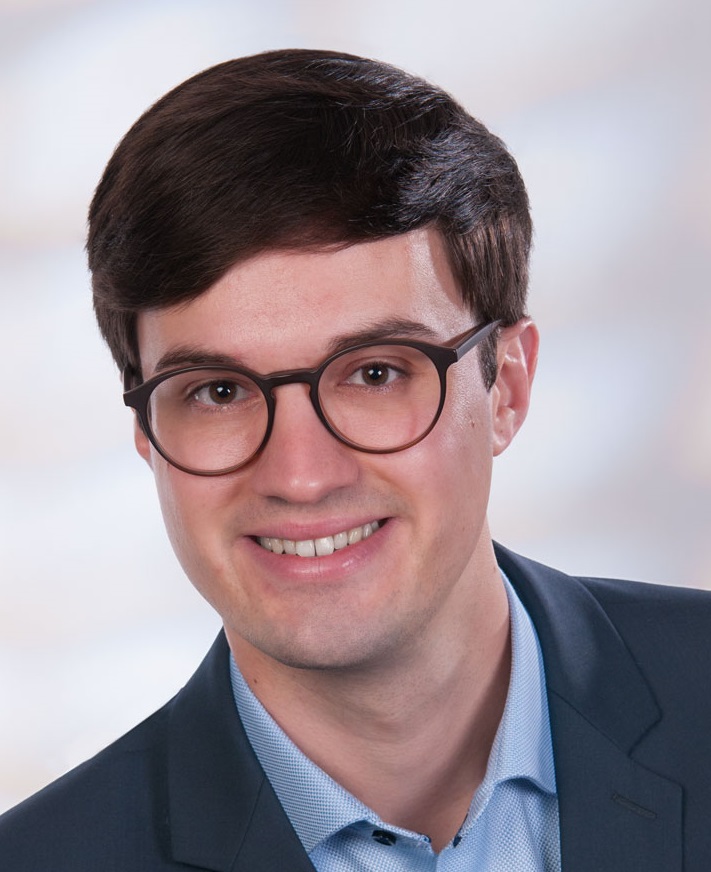}}]{Andreas Venzke} (S'16) received the M.Sc. degree in Energy Science and Technology from ETH Zurich, Zurich, Switzerland in 2017. He is currently working towards the Ph.D. degree at the Department of Electrical Engineering, Technical University of Denmark (DTU), Kongens Lyngby, Denmark. His research interests include power system operation under uncertainty, convex relaxations of optimal power flow and machine learning applications for power systems.
\end{IEEEbiography}
\vspace{-15cm}
\begin{IEEEbiography}[{\includegraphics[width=1in,height=1.25in,clip,keepaspectratio]{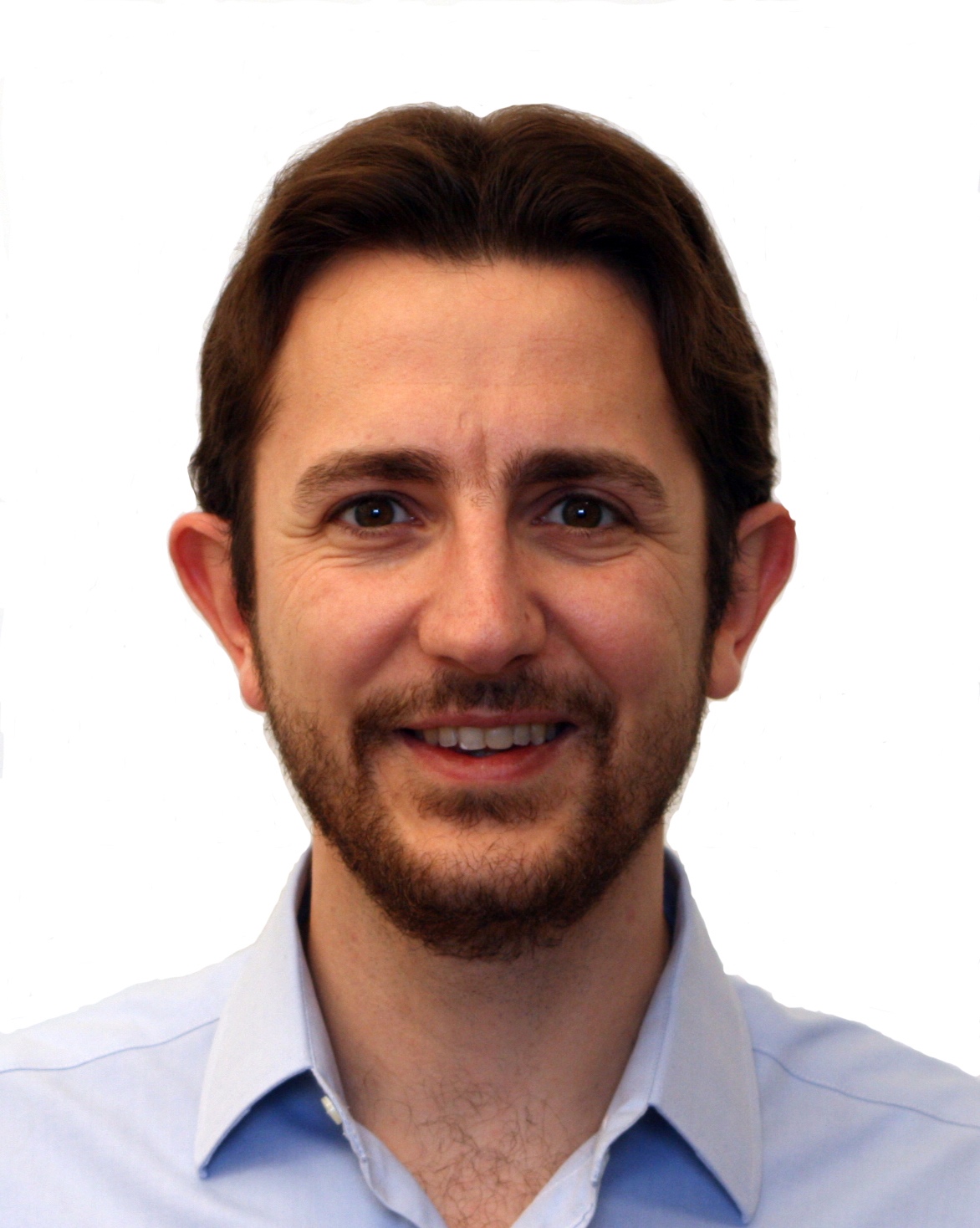}}]{Spyros Chatzivasileiadis} (S'04, M'14, SM'18) is an Associate Professor at the Technical University of Denmark (DTU). Before that he was a post-doctoral researcher at the Massachusetts Institute of Technology (MIT), USA and at Lawrence Berkeley National Laboratory, USA. Spyros holds a PhD from ETH Zurich, Switzerland (2013) and a  Diploma in Electrical and Computer Engineering from the National Technical University of Athens (NTUA), Greece (2007). In March 2016, he joined the  Center for Electric Power and Energy at DTU. He is currently working on power system optimization and control of AC and HVDC grids, and machine learning applications for power systems.
\end{IEEEbiography}

\end{document}